\documentclass[11pt,a4paper]{article}
\pdfoutput=1
\usepackage{jheppub}
\usepackage{amsfonts,amssymb,amsmath}
\usepackage{mathrsfs}
\usepackage{bbm}
\usepackage{youngtab}
\usepackage{rotfloat}
\usepackage{rotating}
\usepackage{tikz}
\usepackage{lscape}

\def\a{\alpha}
\def\b{\beta}
 \def\g{\gamma}

\def\e{\epsilon}

\def\t{\tau}

\def\G{\Gamma}
\def\D{\Delta}

\def\L{\Lambda}

\newcommand{\bC}{\ensuremath{\mathbb{C}}}

\newcommand{\bL}{\ensuremath{\mathbb{L}}}

\newcommand{\bR}{\ensuremath{\mathbb{R}}}

\newcommand{\bV}{\ensuremath{\mathbb{V}}}

\newcommand{\bZ}{\ensuremath{\mathbb{Z}}}

\newcommand{\scD}{\ensuremath{\mathscr{D}}}

\newcommand{\scY}{\ensuremath{\mathscr{Y}}}
\newcommand{\scZ}{\ensuremath{\mathscr{Z}}}

\newcommand{\frakh}{\ensuremath{\mathfrak{h}}}

\newcommand{\frakm}{\ensuremath{\mathfrak{m}}}

\newcommand{\fraksl}{\ensuremath{\mathfrak{sl}}}
\newcommand{\frakgl}{\ensuremath{\mathfrak{gl}}}

\newcommand{\cC}{\mathcal{C}}

\newcommand{\cK}{\mathcal{K}}
\newcommand{\cL}{\mathcal{L}}
\newcommand{\cM}{\mathcal{M}}
\newcommand{\cN}{\mathcal{N}}

\newcommand{\cS}{\mathcal{S}}

\newcommand{\cW}{\mathcal{W}}
\newcommand{\cR}{\mathcal{R}}

\newcommand{\GL}{\mathrm{GL}}
\newcommand{\SU}{\mathrm{SU}}
\newcommand{\SO}{\mathrm{SO}}
\newcommand{\SL}{\mathrm{SL}}

\newcommand{\U}{\mathrm{U}}

\newcommand{\Gr}{{\rm Gr}}
\newcommand{\Fl}{{\rm Fl}}

\newcommand{\Tr}{\mbox{Tr}}

\newcommand{\CP}{{\bf{P}}}

\newcommand{\ind}{{\textrm{ind}}}
\newcommand{\hm}{{\textrm{hm}}}
\newcommand{\vm}{{\textrm{vm}}}
\newcommand{\Hom}{{\textrm{Hom}}}

\def\e{\epsilon}

\def\p{\partial}

\def\bea{\begin{eqnarray}}
\def\eea{\end{eqnarray}}
\def\be{\begin{equation}}
\def\ee{\end{equation}}
\def\ba{\begin{align}}
\def\ea{\end{align}}

\newcommand{\bem}{\begin{pmatrix}}
\newcommand{\eem}{\end{pmatrix}}

\def\={\;  = \;}
\def\+{\, + \,}

\def\wt{\widetilde}
\def\wh{\widehat}
\def\bar{\overline}

\def\rt2{\sqrt{2}}

\def\g{\gamma}
\def\t{\tau}
\def\a{\alpha}
\def\b{\beta}

\def\D{{\Delta}}
\def\L{{\Lambda}}
\def\G{\Gamma}

\preprint{NIKHEF-2014-028}
\title{Givental $J$-functions, Quantum integrable systems, \\ AGT relation with surface operator}

\author{Satoshi Nawata}
\affiliation{NIKHEF theory group,\\ Science Park 105,
1098 XG Amsterdam, The Netherlands \vspace{.2cm}}

\emailAdd{snawata@gmail.com}

\abstract{We study 4d $\cN=2$ gauge theories with a co-dimension two full surface operator, which exhibit a fascinating interplay of supersymmetric gauge theories, equivariant Gromov-Witten theory and geometric representation theory. For pure Yang-Mills and $\cN=2^\ast$ theory, we describe a full surface operator as the 4d gauge theory coupled to a 2d $\cN=(2,2)$ gauge theory. By supersymmetric localizations, we present the exact partition functions of both 4d and 2d theories which satisfy integrable equations. In addition, the form of the structure constants with a semi-degenerate field in $\SL(N,\bR)$ WZNW model is predicted from one-loop determinants of 4d gauge theories with a full surface operator via the AGT relation.}


\begin{document}
\maketitle

\section{Introduction}

In \cite{Gaiotto:2009we}, a large family of 4d $\cN=2$ superconformal field theories (SCFT), known as class $\cS$ theories, has been constructed by compactifying the 6d $\cN=(2,0)$ theory on a Riemann surface. This construction as well as advances in exact results of supersymmetric partition functions has led to the celebrated AGT relation \cite{Alday:2009aq,Wyllard:2009hg}, which amounts to the statement that the partition function of a 4d $\cN = 2$ SCFT on $S_b^4$ \cite{Pestun:2007rz,Hama:2012bg} can be identified with a correlation function of 2d Toda CFT on the corresponding Riemann surface. 

The AGT relation becomes particularly enriched when we insert a half-BPS non-local operator called a surface operator \cite{Gukov:2006jk} supported on $S^2\subset S_b^4$. One can characterize a surface operator by specifying the boundary condition of the gauge field on $S^2\subset S_b^4$ which breaks the gauge group to the Levi subgroup $\bL\subset G$. In this paper, we consider the $\SU(N)$ gauge group so that  the Levi subgroup $\bL=\textrm{S}[\U(N_1)\times\cdots\times\U(N_M)]$ is specified by a partition $N=N_1+\cdots+N_M$ which we denote $[N_1,\cdots,N_M]$. Especially, the surface operator of $[1,N-1]$-type is called simple and  that of $[1,\cdots,1]$-type is denoted full. Moreover, the dynamics on a surface operator is described by coupling a 2d gauge theory to the 4d bulk theory  \cite{Dimofte:2010tz,Gadde:2013dda,Gaiotto:2013sma,Gomis:2014eya}.

From the 6d view point, there are two ways to realize a surface operator. One way is to attach a collection of M2-branes on the M5-branes and we call it a co-dimension four surface operator. 
It was argued in \cite{Alday:2009fs} that the insertion of a completely degenerate field in Toda CFT corresponds to a co-dimension four simple surface operator in a 4d gauge theory via the AGT relation. Thus, the Nekrasov partition function with the surface operator satisfies the BPZ equation \cite{Belavin:1984vu}. Recently, the authors of \cite{Gomis:2014eya} provide a complete microscopic description of a general co-dimension four surface operator in terms of a 2d $\cN=(2,2)$ gauge theory coupled to the 4d $\cN=2$ gauge theory and identify the corresponding degenerate operator in Toda CFT labelled by a Young diagram.

On the other hand, the intersection of M5-branes spanning $S^2\subset S_b^4$ and wrapping a Riemann surface also gives rise to a surface operator in the 4d gauge theory, which we denote a co-dimension two surface operator. The effect of wrapping the defect on the Riemann surface results in the change of the symmetry in 2d CFT. For a surface operator of type $\vec{N}=[N_1,\cdots,N_M]$, it was conjectured \cite{Braverman:2010ef,Wyllard:2010rp} that the 2d symmetry is the W-algebra $W(\wh\fraksl(N),\vec{N})$ obtained by the quantum Drinfeld-Sokolov reduction \cite{Bershadsky:1989mf,Feigin:1990pn,deBoer:1993iz} for the embedding $\rho_{\vec{N}}:\fraksl(2)\to\fraksl(N)$ corresponding to the partition $\vec{N}$. For the 4d gauge theory side, the moduli space of instanton with the boundary condition of the gauge field on the surface is called \emph{affine Laumon space}. It was shown in \cite{Feigin:2008,Finkelberg:2010} that  the affine Laumon space is equivalent to instanton moduli space on an orbifold $\bC\times (\bC/\bZ_M)$ so that it admits quiver representations, called \emph{chain-saw quivers}.  Using the quiver representations, one can demonstrate localization computations of the Nekrasov partition functions \cite{Feigin:2008,Kanno:2011fw}. It was checked in \cite{Wyllard:2010rp,Wyllard:2010vi,Tachikawa:2011dz,Kanno:2011fw} \cite[\S6.1]{Tan:2013tq} that  the instanton partition function of the pure Yang-Mills theory with a surface operator of type $\vec{N}$ is equal to the norm of the Gaiotto-Whittaker state in the Verma module of the W-algebra $W(\wh\fraksl(N),\vec{N})$. In particular, for a full surface operator $[1,\ldots,1]$, more extensive checks have been carried out \cite{Alday:2010vg,Kozcaz:2010yp,Awata:2010bz} for the correspondence between  instanton partition functions and conformal blocks of the affine Lie algebra $\wh\fraksl(N)$. In this paper, we shall provide the contour integral expressions of the Nekrasov partitions functions for the pure Yang-Mills and the $\cN=2^*$ theory with a surface operator by using the supersymmetric non-linear sigma model with the chain-saw quiver as a target.

The Nekrasov partition functions in the presence of a surface operator encode both 4d and 2d non-perturbative dynamics. Hence, when we turn off the 4d instanton effect, the Nekrasov partition functions reduce to 2d vortex partition functions which contains the non-perturbative dynamics on the support of the surface operator. In fact, when the instanton number is zero, the chain-saw quivers demote to hand-saw quivers so that the generating function of equivariant cohomology of the hand-saw quivers becomes the vortex partition function. On the other hand, a surface operator can also be described as a coupling of the 4d gauge theory with a 2d theory on the surface. In particular, the description on a surface operator in the pure Yang-Mills is given by a coupling of the $\cN=(2,2)$ non-linear sigma model with a flag manifold $G/\bL$. In addition, for the $\cN=2^*$ theory,  the $\cN=(2,2)^*$ non-linear sigma model with the cotangent bundle $T^*(G/\bL)$ of the flag manifold depicts the dynamics on the support of the surface operator. Since their ultra-violet descriptions as $\cN=(2,2)$ gauged linear sigma models are known, one can also compute the vortex partition functions by means of Higgs branch localizations \cite{Benini:2012ui,Doroud:2012xw}. Therefore, we will see the correspondence of vortex partition functions computed by the two methods.

In this paper, we will also demonstrate explicit calculations for one-loop determinants when a full surface operator is inserted. The $\cN=2$ partition functions on $S^4_b$ require both the Nekrasov partition functions and one-loop determinants over the instanton configurations \cite{Pestun:2007rz,Hama:2012bg}. Since the Nekrasov partition functions in the presence of a full surface operator can be computed by the orbifold method, it is plausible to expect that the one-loop determinants with a full surface operator is equivalent to those on the orbifold space $\bC\times (\bC/\bZ_N)$. In fact, we show that the one-loop determinants calculated by using the index theory on $\bC\times (\bC/\bZ_N)$ correctly encode both 4d and 2d perturbative contributions. 

In the AGT relation, the Nekrasov partition functions correspond to the conformal blocks while the one-loop determinants is equivalent to the product of the three-point functions of 2d CFT. When $N=2$, the one-loop determinants of 4d gauge theories with a full surface operator computed by the orbifold procedure reproduce the structure constant of $\SL(2,\bR)$ WZNW model determined in \cite{Teschner:1997ft,Teschner:1999ug,Maldacena:2001km}. Furthermore, using the one-loop determinants of 4d gauge theories with a full surface operator, we predict the form of the two-point and three-point functions of $\SL(N,\bR)$ WZNW model.

Let us also mention the algebro-geometric aspect of the AGT relation with a surface operator. The fundamental idea of algebraic topology is to extract algebraic objects which encode the information of a given space.
Homology, cohomology groups and fundamental groups can be seen as typical examples for this idea. This idea has resulted in a great success in mathematics of the 20th century. From the late 80s, inspired by the idea coming from quantum field theory and string theory, ``quantizations'' of these invariants in algebraic topology have been introduced, which opened up to the dawn of new geometry and quantum topology. In particular, one of significant steps to uncover deeper structures behind ``quantization'' has been made by Givental \cite{Givental:1994,Givental:1995s,Givental:1996}. Since Givental's theory plays an essential role in this paper, let us briefly review it by using a projective space $\CP^{N-1}$ as an example. 

It is well-known that the cohomology ring of $\CP^{N-1}$ is isomorphic to
\bea
H^*(\CP^{N-1})\cong \bC[x]/(x^N)~.
\eea
The cohomology ring relation $x^N=0$ can be resolved by using equivariant cohomology. To see that explicitly, let us define the $S^1$-equivariant action on  $\CP^{N-1}$ by
\be
\lambda [z_0:\cdots:z_{N-1}]=[\lambda^{r_0}z_0:\cdots:\lambda^{r_{N-1}}z_{N-1}]~,
\ee
for $\lambda\in S^1$. Then, the  $S^1$-equivariant cohomology ring of $\CP^{N-1}$ is given by
\bea
H^*_{S^1}(\CP^{N-1})\cong \bC[x,\hbar]/(\prod_{i=0}^{N-1} (x-r_i\hbar))~,
\eea
where $\hbar$ represents the hyperplane class of the base manifold of the universal $S^1$-bundle
\be
S^{2\infty+1}=ES^1\to BS^1=\CP^{\infty}~,
\ee
so that $H^*(BS^1)=H^*(\CP^{\infty})=\bC[\hbar]$. Here the hyperplane class $\hbar$  plays a similar role to the Planck constant so that it resolves the cohomology ring relation. Moreover, the cohomology ring is quantized based on Gromov-Witten theory. The quantum cohomology is ordinary cohomology with a quantum product defined by
\bea
T_i \circ T_j=\sum_{k,\ell} C_{ijk}(t)\eta^{k\ell} T_\ell~,
\eea
for a basis $T_i$ of the cohomology group. Here the structure constants $C_{ijk}(t) :=\frac{\partial^3 F_0}{\p T_i \p T_j \p T_k}$ is the third derivative of the genus-zero prepotential depending on the complexified K\"ahler parameter $t$ and $\eta_{ij}:=\int T_i \cup T_j$ is the metric on the cohomology group. In fact, the WDVV equation is equivalent to the associativity of the quantum product, and therefore the quantum product can be thought of as quantum deformation of the cup product of cohomology. Writing $q=e^{t}$, the quantum cohomology ring of $\CP^{N-1}$ is isomorphic to
\bea
QH^*(\CP^{N-1})\cong \bC[x,q]/ (x^N-q)~.
\eea
One of the most intriguing aspects of quantum cohomology is its relation with differential equations. Actually, Givental's profound insight perceived the relation in the equivariant Floer homology of the loop space. Roughly speaking, the Floer homology is the $\frac{\infty}{2}$-dimensional homology theory of infinite-dimensional manifolds. In this example, it is suitable to consider the universal covering $\wt{L\CP^{N-1}}$ of the loop space  ${L\CP^{N-1}}:=\textrm{Map}(S^1,\CP^{N-1})$  of the projective space. For this space, one can obtain the explicit expression of the $S^1$-equivariant Floer homology 
\bea
HF^*_{S^1}(\wt{L\CP^{N-1}})=\bigoplus_{m\in \bZ} \bigoplus_{k=0}^{N-1} \bC[\hbar]\cdot (x-m\hbar)^k\cdot \prod_{j<m} (x-j\hbar)^N~.
\eea
Remarkably, the $S^1$-equivariant Floer homology $HF^*_{S^1}(\wt{L\CP^{N-1}})$ turns out to be endowed with $\scD$-module structure 
\bea\label{d-module}
\scD/(p^N-q)~,
\eea
where we define
\bea
p\cdot J(x,\hbar)&=&x\cdot J(x,\hbar)~,\cr
q\cdot J(x,\hbar)&=& J(x-\hbar,\hbar)~,
\eea
for $J(x,\hbar)\in  HF^*_{S^1}(\wt{L\CP^{N-1}})$. From the definition, it is easy to see $[p,q]=\hbar q$ so that $p$ can be regarded as a differential operator $\hbar q \frac{d }{dq}$ on functions of $q$. Therefore, the $\scD$-module structure \eqref{d-module} can be rephrased as
\bea\label{quantum-connection}
\Bigg[\Big(\hbar q\frac{d}{dq}\Big)^N-q\Bigg]J(q)=0~.
\eea
Usually, this differential equation is called a quantum (Dubrovin) connection, which can be considered as a ``quantum curve" of the quantum cohomology ring. This directly leads to the theory of integrable systems because $\scD$-modules of this kind can be written as flat connections. Furthermore, it turns out that the solution of the quantum connection \eqref{quantum-connection} is given by the generating function of the equivariant genus-zero  Gromow-Witten invaraints with gravitational descendants which is called Givental's $J$-function of $\CP^{N-1}$
\bea
J[\CP^{N-1}]=e^{\frac{tx}{\hbar}}\sum_{d=0}^{\infty}\frac{e^{td}}{\prod_{j=1}^d(x+j\hbar)^N}~.
\eea
The precise definition of the $J$-function of a compact K\"ahler variety is given in \S\ref{sec:flag}. 

The $J$-functions are sublated by Braverman and Etingof to geometric representation theory \cite{Braverman:2004vv,Braverman:2004cr}. In \cite{Braverman:2004vv}, the invariant equivalent to the $J$-function of the complete flag variety $\Fl_N$ has been constructed as the generating function of the equivaraint cohomology of the moduli space of quasi-maps $\CP^1\to \Fl_N$. The moduli space of quasi-maps $\CP^1\to \Fl_N$ is called \emph{Laumon space} which is indeed described by the hand-saw quivers. Strikingly, the equivaraint cohomology of the Laumon space turns out to be isomorphic to the Verma module of the Lie algebra $\fraksl(N)$. Moreover, this relation can be uplifted to the infinite-dimensional version by using the affine complete flag variety which can be thought of as a complete flag variety for the loop group. In fact, the moduli space of quasi-maps from $\CP^1$ to the affine complete flag variety is the affine Laumon space, and its equivariant cohomology receives the action of the affine Lie algebra $\wh \fraksl(N)$. Therefore, this can be naturally interpreted in the context of the AGT relation with a full surface operator. From this view point, the Nekrasov partition function of the pure Yang-Mills  with a full surface operator can be considered as the $J$-function of the affine complete flag variety. 
In addition, the geometric representation theoretic aspect of the $\cN=2^*$ theory with a full surface operator has been studied by Negut \cite{Negut:2008,Negut:2011}. In this paper, we just provide a physical interpretation of the results in \cite{Braverman:2004vv,Braverman:2004cr,Negut:2008,Negut:2011}. Nevertheless, the AGT relation of class $\cS$ theories with a surface operator generally provides a rich arena for a vast generalization of Givental theory, and quantum connections therefore appear as differential equations of Knizhnik-Zamolodchikov type.

The paper is outlined as follows. In \S \ref{sec:gauge}, we provide a microscopic description of a full surface operator and give explicit formulae of the partition functions of 4d and 2d gauge theory for the pure Yang-Mills and the $\cN=2^\ast$ theory. Most of the results in this section have already been proven in literature of mathematics \cite{Givental:1995,Givental:2003,Bertram:2004,Braverman:2004vv,Braverman:2004cr,Braverman:2006,Negut:2008,Finkelberg:2010,Braverman:2010,Negut:2011}. What is mathematically new is that we conjecture the explicit expression of the $J$-function of the cotangent bundle of the complete flag variety by using the supersymmetric partition function on $S^2$. In addition, we show the evidence that the one-loop determinants can be computed by the orbifold method. In \S  \ref{sec:3pt}, we predict the form of the two-point and three-point function of $\SL(N,\bR)$ WZNW model by using the one-loop determinants of the 4d gauge theory. \S \ref{sec:discussion} is devoted to discuss future directions. In Appendix \ref{sec:instanton}, we derive the contour integral expressions of  Nekrasov partition functions with a general surface operator, and  computations of one-loop determinants by means of the Atiyah-Singer index theory is given in Appendix \ref{sec:1-loop}. Finally, the $J$-function of the cotangent bundle of a partial flag variety is presented in Appendix \ref{sec:partial}.

\section{Gauge theory with full surface operator}\label{sec:gauge}

The surface operator was first introduced as a half-BPS non-local operator supported on a surface in the $\cN=4$ SCFT by Gukov and Witten \cite{Gukov:2006jk}. One way to define a surface operator is to specify a singular behavior of gauge fields on the surface. To describe more precisely, let $(z_1,z_2)$ be complex coordinate and the surface operator is supported on the plane $\cC=\{(z_1,z_2)|z_2=0\}$. If $(r,\theta)$ is the polar coordinate of $z_2$-plane, the singular behavior of gauge fields is prescribed as
\bea\label{singular}
A_\mu dx^\mu\sim {\rm diag}(\alpha_1,\ldots,\alpha_N) id\theta~,
\eea
on the place $\cC$. Thus, the parameters $\vec{\a}=(\alpha_1,\ldots,\alpha_N)$ can be considered as the monodromies of the abelian gauge fields around the operator.  If the singular date has the structure \begin{equation}
\vec\alpha=(
\underbrace{\alpha_{(1)},\ldots,\alpha_{(1)}}_{\text{$N_1$ times}},
\underbrace{\alpha_{(2)},\ldots,\alpha_{(2)}}_{\text{$N_2$ times}},
\ldots,
\underbrace{\alpha_{(M)},\ldots,\alpha_{(M)}}_{\text{$N_M$ times}} )~,\label{ordered}
\end{equation}
where $\alpha_{(I)}>\alpha_{(I+1)}$, the gauge group is broken to the commutant of $\vec{\a}$ on the surface $\cC$:
 \begin{equation}
\bL=\textrm{S}[\U(N_1)\times \U(N_2)\times \cdots \times \U(N_M)]~,
\end{equation} 
which is called the Levi subgroup. In fact, the subgroup $\bL$ is the Levi part of a parabolic subgroup $\mathcal{P}$ of the complexified Lie group $G_\bC$. For instance, if $\vec{\alpha}=(\alpha, \cdots, \alpha, (1-N)\alpha)$, the Levi group is $\bL=\SU(N-1)\times \U(1)$, which is called simple.  When all $\alpha_i$ are distinct,  which is called a full surface operator,  the Levi group is $\bL=\U(1)^N$ and the corresponding parabolic group is the Borel subgroup $B$ of $\SL(N,\bC)$. In addition to the $M$ continuous parameters $\alpha_{(I)}$, there are ``electric parameters'' or ``2d theta angles'' $\eta_I$ corresponding to $\U(1)^{M}\in \bL$. These parameters enter into the path integral through the phase factor $\exp(i\eta_I \frakm^I )$ where $\frakm^I$ are magnetic fluxes on $\cC$ 
\bea
\frakm^I=\frac{1}{2\pi} \int_\cC F^I \qquad (I=1,\cdots,M)~,
\eea
where $\sum_I\mathfrak{m}^{I}=0$. 

The other way to describe a surface operator is to couple a 4d $\cN=2$ gauge theory to an $\cN=(2,2)$ supersymmetric gauge theory on the surface $\cC$ \cite{Gukov:2006jk}. For the surface operator in the pure Yang-Mills, the 2d theory flows at infrared to the  $\cN=(2,2)$ supersymmetric non-linear sigma model (NLSM) with the partial flag variety $G_\bC/\mathcal{P}$ as a target. In this description, the combined parameters $\vec{t}=2\pi i(\vec{\eta}+i\vec{\alpha})$ are identified with the complexified K\"ahler parameters of the NLSM. Furthermore, for the surface operator in the $\cN=2^*$ theory, the infrared description of the 2d theory is given by an $\cN=(2,2)$ NLSM with the cotangent bundle $T^*(G_\bC/\mathcal{P})$ of the flag variety. 

\begin{figure}[htbp]
\begin{center}
 \includegraphics[width=14cm]{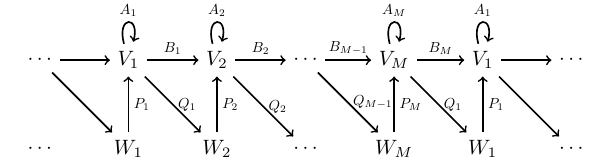}
\caption{Chain-saw quiver}
\label{fig:chain-saw}
\end{center}
\end{figure}

The instanton configurations $F=-\ast \!F$ on $\bR^4\backslash \cC$ with the singularity \eqref{singular} are called ramified instantons. The moduli space of the ramified instantons is characterized by the Levi subgroup with $\vec{N}=[N_1,\cdots,N_M]$, the instanton number $k$ and the magnetic fluxes $\frakm^I$ so that we denote it by $\cM_{\vec{N},k,\vec{\frakm}}$.  The corresponding objects in algebraic geometry is actually rank-$N$ torsion-free sheaves on $\CP^1\times \CP^1$ with coordinates $(z_1,z_2)$, with framing given at $\{z_1=\infty\}\cup\{z_2=\infty\}$ and with parabolic structure of type $\mathcal{P}$ given at $\{z_2=0\}$, called affine Laumon space \cite{Feigin:2008,Finkelberg:2010}. The affine Laumon space can be also regarded as the smooth resolution of the space of quasi-maps from $\CP^1$ into affine flag variety \cite{Braverman:2010}. Furthermore, using the equivalence between a parabolic sheaf on $\CP^1\times \CP^1$ of type $\mathcal{P}$ and a $\bZ_M$-equivariant sheaf on $\CP^1\times \CP^1$, the quiver description of $\cM_{\vec{N},k,\vec{\frakm}}$ is given by the ADHM quiver on the orbifold space $\bC\times (\bC/\bZ_M)$. The resulting quiver is called a chain-saw quiver shown in Figure \ref{fig:chain-saw}.  In this prescription, it is convenient to combine the instanton number $k$ and the magnetic fluxes $\frakm^I$ as follows:
\begin{equation}\label{k-m}
k_M=k,\qquad k_{I+1}=k_I+\mathfrak{m}^{I+1}~,
\end{equation} 
where the index $I$ is taken modulo $M$. Thus, we also denote the moduli space of ramified instantons by $\cM_{\vec{N},\vec{k}}$ with $\vec{k}=[k_1,\cdots,k_M]$. To describe the ADHM construction of $\cM_{\vec{N},\vec{k}}$, let $V_I$ and $W_I$ $(I=1,\cdots,M)$ vector spaces of dimension
\be
\dim W_I = N_I, \qquad  \dim V_I = k_I~,
\ee
and we denote
$A_{I } \in \mathrm{Hom}~(V_I ,V_I )$,  $B_I  \in \mathrm{Hom}~(V_{I },V_{I +1})$,
$P_{I } \in \mathrm{Hom}~(W_I ,V_I )$ 
and $Q_{I } \in \mathrm{Hom}~(V_I ,W_{I +1})$.
Then, the ADHM equations are
\be\label{ADHM-eq}
\mathcal{E}^{(I)}_{\bC}:=A_{I +1} B_{I } - B_{I } A_{I } + P_{I +1} Q_{I } = 0~.
\ee
where the index $I$ is taken modulo $M$. The moduli space is given by
\be
\cM_{\vec{N},\vec{k}}=\{(A_{I } ,B_I,P_I,Q_I) |\mathcal{E}^{(I)}_{\bC}=0,~ \textrm{stability condition}\}/\GL(k_1,\bC)\otimes\cdots\otimes\GL(k_M,\bC)~.
\ee
As in the case without a surface operator, the moduli space $\cM_{\vec{N},\vec{k}}$ of ramified instantons receives the action of the Cartan torus  $\U(1)^2\times \U(1)^N$ of the spacetime and the gauge symmetry. Due the the orbifold operation, one of the equivariant parameters of $\U(1)^2$ acts on the spacetime coordinate fractionally as
\be\label{equiv-quotient}
(z_1,z_2)\to (e^{i\e_1} z_1,e^{i\e_2/M}z_2)~.
\ee
In addition, since there are non-contractible cycles in the asymptotic region of $\bC\times (\bC/\bZ_M)$, the gauge field can have a non-trivial holonomy. The non-trivial holonomy shifts the equivariant parameters $(a_1,\cdots,a_N)$ of $\U(1)^N$ by
\bea\label{holonomy-shift}
a_{s,I} \to a_{s,I}-\tfrac{I-1}{M}\e_2~, \qquad (s=1,\cdots,N_I) ~.
\eea
Fixed points under the equivariant action can be labeld by  $\vec{N}$-tuple of Young diagrams. For more detail, we refer the reader to \cite{Kanno:2011fw}. Subsequently, the character of the equivariant action at the fixed points yields the Nekrasov insanton partition function \cite{Feigin:2008,Kanno:2011fw}
\bea\label{inst1}
\scZ_{\mathrm{inst}}[\vec{N}] = \sum_{\vec{k}} \prod_{I=1}^M z_I^{k_I} \scZ_{\vec{N},\vec{k}}(\e_1,\e_2,a,m)~,
\eea
where $z_I$ are instanton counting fugacity and $Z_{\vec{N},\vec{k}}(\e_1,\e_2,a,m)$ depends on the matter content of the theory. In the context of the AGT relation, it was conjectured \cite{Braverman:2010ef} that the 2d symmetry is the W-algebra $W(\wh\fraksl(N),\vec{N})$ obtained by the quantum Drinfeld-Sokolov reduction \cite{Bershadsky:1989mf,Feigin:1990pn,deBoer:1993iz} for the embedding $\rho_{\vec{N}}:\fraksl(2)\to\fraksl(N)$ corresponding to the partition $\vec{N}$. In particular, when a full surface operator is present, it wan first proven in \cite{Braverman:2004vv,Braverman:2004cr} that the equivariant cohomology of the ramified instanton moduli space $\cM_{[1^N],\vec{k}}$ receives the action of the affine Lie algebra $\wh\fraksl(N)$. The checks of the correspondence between instanton partition functions of 4d SCFTs and $\wh\fraksl(N)$ conformal blocks have been carried out in \cite{Alday:2010vg,Kozcaz:2010yp,Awata:2010bz}. For general $W$-algebras, it has been checked in \cite{Wyllard:2010rp,Wyllard:2010vi,Kanno:2011fw} that the ramified instanton partition functions of the pure Yang-Mills match with the norm of the Gaiotto-Whittaker states in the Verma module of the corresponding $W$-algebra.

\begin{figure}[htbp]
\begin{center}
 \includegraphics[width=12cm]{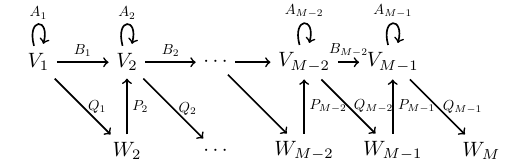}
\caption{Hand-saw quiver}
\label{fig:hand-saw}
\end{center}
\end{figure}

By making change of variables
\be\label{cov}
 z_I=e^{t_I-t_{I+1}} ~ \ (I=1,\cdots,M-1)~ ~,\qquad \prod_{I=1}^M z_I=q~,
\ee
the instanton partition function \eqref{inst1} can be re-arranged with \eqref{k-m} as
\be
\scZ_{\mathrm{inst}}[\vec{N}] = \sum_{k=0}^\infty \sum_{\mathfrak{m}\in \Lambda_\mathbb{L} }
q^k e^{t\cdot \mathfrak{m}} \scZ_{\vec{N},k,\vec{\frakm}}(\e_1,\e_2,a,m)~,
\ee
If a theory is superconformal, the fugacity of the instanton number $k$ can be expressed in terms of the complexified gauge coupling $\tau$ by $q = e^{2 \pi i \tau}$.
For an asymptotically free theory, it is replaced by the dynamical scale $\Lambda$ with appropriate mass dimension. 
The chemical potentials $\vec{t}$ for the magnetic fluxes $\vec{\frakm}$ are indeed the 2d complexified K\"ahler parameters  $\vec{t}=2\pi i(\vec{\eta}+i\vec{\alpha})$. Hence, when the instanton number is zero $k=0$, the partition function encodes only 2d dynamics on the support of the surface operator. Moreover, the $k=0$ specialization of the chan-saw quiver in Figure \ref{fig:chain-saw} reduces to the hand-saw quiver \cite{Finkelberg:2010} in Figure \ref{fig:hand-saw}, which is equivalent to the smooth resolution of the space of quasi-maps from $\CP^1$ into the flag variety, called \emph{Laumon space} \cite{Nakajima:2012}. The finite $W$-algebra that can be obtained by quantum Drinfeld-Sokolov reduction of Lie algebra acts on the equivariant cohomology of the Laumon space \cite{Nakajima:2012}. We shall show that the generating function of the equivariant cohomology of the Laumon space is actually the vortex partition function of the $\cN=(2,2)$ NLSM with the partial flag variety specified by the partition $\vec{N}$. In particular,  the generating function of the equivariant cohomology of the Laumon space can be identified with the Givental $J$-function of the flag variety \cite{Braverman:2004vv}.

In this section, we concentrate on the pure Yang-Mills and the $\cN=2^*$ theory with a full surface operator. For these theories, the Nekrasov partition functions and the vortex partition functions on the support of the surface operator obey differential equations. Since they can be interpreted as quantum connections of Givental $J$-functions, they are written as integrable Hamiltonians. The pure Yang-Mills is related to the Toda integrable system  \cite{Givental:1995,Givental:2003,Braverman:2004vv,Braverman:2004cr} whereas  the $\cN=2^*$ theory is connected to the Calogero-Moser integrable system \cite{Negut:2008,Braverman:2010,Negut:2011}. When a general surface operator is placed, we present the partition functions in Appendix \ref{sec:instanton} and \ref{sec:partial}.

Since an $\cN=2$ supersymmetric path integral on $S_b^4$ localizes on the (anti-)instanton configurations on the north (south) pole,
in order to obtain full exact partition functions on $S_b^4$, one-loop determinants over the (anti-)instanton configurations have to be computed in addition to instanton partition functions \cite{Pestun:2007rz,Hama:2012bg}. When a surface operator is present, the calculations of one-loop determinants  have not been demonstrated although the literature \cite{Feigin:2008,Alday:2010vg,Kozcaz:2010yp,Awata:2010bz,Wyllard:2010rp,Wyllard:2010vi,Kanno:2011fw} has evaluated instanton partition functions. As in the case of instanton partition functions, it is natural to expect that the one-loop determinants can be evaluated by the orbifold method. In this paper, we propose that one-loop determinants in the existence of a full surface operator can be obtained by means of the Atiyah-Singer index theorem for transversally elliptic operators on the orbifold space $\bC\times (\bC/\bZ_N)$. 
 To support this statement, we shall show that the one-loop determinants computed by this method correctly contain both the 4d and 2d perturbative contributions.

\subsection{Pure Yang-Mills}
\subsubsection{Instanton partition function}
The pure $\SU(N)$  Yang-Mills theory is obtained by wrapping $N$ M5-branes on a two-punctured sphere. Although the instanton partition function of the pure Yang-Mills with a surface operator is expressed as a character of the equivariant action at the fixed points of the chain-saw quiver shown in Figure \ref{fig:chain-saw} \cite{Kanno:2011fw}, here we yield the contour integral representation of the $\U(N)$ instanton partition function by using the supersymmetric NLSM with the chain-saw quiver as a target. Since the detail is presented in Appendix \ref{sec:instanton}, we just give the expression \eqref{pure-inst-so} for the $\U(N)$ instanton partition function of the $\cN=2$ pure Yang-Mills theory with a full surface operator
\bea
\scZ_{\textrm{inst}}^{\textrm{pure}}[1^N]=\sum_{\vec{k}}  \Big( \prod_{I=1}^N z_I^{k_I} \Big) \scZ_{[1^N],\vec{k}}^{\textrm{pure}}~,
\eea
where
\bea
\scZ_{[1^N],\vec{k}}^{\textrm{pure}} &=&\epsilon_1^{-\sum_{I=1}^Nk_I} \oint  \prod_{I=1}^N \prod_{s=1}^{k_I} \dfrac{d \phi_s^{(I)}}{ (\phi_s^{(I)} + a_I-\frac{(I-1)\e_2}{N}) (\phi_s^{(I)} +a_{I+1}+\e-\frac{I\e_2}{N})} \cr
&& \hspace{2cm}\prod_{I=1}^N \prod_{s=1}^{k_I} \prod_{t \neq s}^{k_I} \dfrac{ \phi_{st}^{(I)} }{\phi_{st}^{(I)} + \epsilon_1} 
\prod_{I=1}^N \prod_{s=1}^{k_I} \prod_{t=1}^{k_{I+1}} \dfrac{\phi_{s}^{(I)} - \phi_t^{(I+1)} + \epsilon}{\phi_{s}^{(I)} - \phi_t^{(I+1)} + \frac{\epsilon_2}{N}}~.
\eea
 The $\SU(N)$ instanton partition function could be obtained by simply dropping the ``$\U(1)$ factor" \cite{Alday:2009aq,Kanno:2011fw}. Then, the $\SU(N)$ instanton partition function is dual to the norm of a coherent state, called the Gaiotto-Whittaker state, of the Verma module of the affine Lie algebra $\wh \fraksl(N)$ \cite{Kozcaz:2010yp,Kanno:2011fw}. Interestingly,
the instanton partition function satisfies the periodic Toda equation \cite{Yamada:2010rr}
\begin{equation}
\left[\frac{\epsilon_1^{2}}{2} \sum_{I=1}^N (z_I\partial_I-z_{I+1}\partial_{I+1})^2+\epsilon_1\sum_{I=1}^N u_I z_I\partial_I-\sum_{I=1}^N z_I\right]{\scZ}_{\textrm{inst}}^{\textrm{pure}}[1^N] =0~,
\end{equation}
where we impose  the periodic condition $z_{N+I}=z_I$ on $z$ and 
\bea
u_I=a_{I+1}-a_I~,\qquad u_{I+N}=u_I+\e_2~.
\eea
In fact, making the change of variables as in \eqref{cov}
\bea\label{cov2}
  z_I=e^{t_{I}-t_{I+1}} ~ \ (I=1,\cdots,N-1)~,\qquad \prod_{I=1}^N z_I=\Lambda ~,
\eea
where $\Lambda$ can be interpreted as the dynamical scale of the pure Yang-Mills, one can bring the equation into the more familiar form
\be\label{periodic-toda}
\left[2\e_1\e_2\L \frac{\p}{\p\L}+\e_1^2\Delta_{\frakh} -2\Big(\L e^{t_N-t_1}+\sum_{\alpha\in \Pi} e^{\langle t,\alpha\rangle} \Big)\right](e^{-\frac{\langle a,t\rangle}{\e_1}}{\scZ}_{\textrm{inst}}^{\textrm{pure}}[1^N]  )={\langle a,a\rangle}(e^{-\frac{\langle a,t\rangle}{\e_1}}{\scZ}_{\textrm{inst}}^{\textrm{pure}}[1^N]  )~,
\ee
where $\Pi$ represents the set of simple roots of $\fraksl(N)$ so that $\sum_{\alpha\in \Pi} e^{\langle t,\alpha\rangle}=\sum_{I=1}^{N-1} e^{t_{I}-t_{I+1}}$, and the rest of notations is as follows:
\bea
\Delta_{\frakh}=\sum_{I=1}^{N}\frac{\partial^2}{\partial t_I^2} ~, \qquad
 \langle a ,  t\rangle=\sum_{I=1}^{N} a_{I}t_{I}~,\qquad \langle a,a\rangle= \sum_{I=1}^Na_{I}^2~.
\eea
This was first derived in the context of geometric representation theory \cite{Braverman:2004vv,Braverman:2004cr} and later reproduced in the context of the AGT relation \cite{Awata:2009ur,Yamada:2010rr}.

\subsubsection{$J$-function of complete flag variety}\label{sec:flag}

\begin{figure}[h]
 \centering
 \includegraphics[width=10cm]{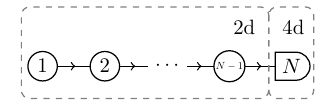}
    \caption{Quiver diagram of the 2d-4d coupled system for the pure Yang-Mills in the presence of a full surface operator where we use the hybrid node as in \cite{Gomis:2014eya}  to denote a 4d gauge group which gauges a 2d flavor symmetry. The Higgs branch of $\cN=(2,2)$ GLSM is the complete flag variety.}\label{fig:flag}
\end{figure}

Since the surface operator is a half-BPS operator, it preserves four supercharges. Moreover, the surface operator can be also described as a 2d $\cN=(2,2)$ supersymmetric gauge theory coupled to the 4d $\cN=2$ gauge theory. For a full surface operator in the 4d $\cN=2$ pure Yang-Mills, the 2d $\cN=(2,2)$ supersymmetric gauge theory coupled to the 4d pure Yang-Mills is described by the quiver diagram above (Figure \ref{fig:flag}). At UV, the matter content consists of bifundamentals $(\bf 1,\bar{2})\oplus \ldots \oplus (N-2,\bar{N-1})$ and $N$ fundamentals $\bf N-1$. The 2d quiver gauge theory is coupled to the 4d pure Yang-Mills  by gauging the flavor symmetry $\U(N)$. Hence, the Coulomb branch parameters $a_i$ in the 4d theory become the twisted masses of the fundamentals in the 2d theory. Since the Higgs branch of the 2d theory is given by the complete flag variety $\Fl_N=\SL(N,\bC)/B$ where $B$ is the Borel subgroup of $\SL(N,\bC)$, the 2d theory flows to the NLSM with the complete flag variety $\Fl_N$ in the infrared. It is worth mentioning that there is another description for the complete flag variety as an increasing sequence of linear subspaces of $\bC^N$
\bea
0 \subset \bC \subset \bC^2 \subset \cdots \subset \bC^{N-1} \subset \bC^N~,
\eea
which indeed yields the quiver description. In fact, the description by the gauged linear sigma model (GLSM) presented in Figure \ref{fig:flag} enables us to compute the exact partition function of the $\cN=(2,2)$ quiver gauge theory on $S^2$ \cite{Benini:2012ui,Doroud:2012xw}. From the $S^2$ partition functions, one can extract the Givental $J$-function of the Higgs branch of the GLSM \cite{Bonelli:2013mma}, which plays an important role in this paper.

Therefore, let us briefly recall the definition of the Givental $J$-function of a compact K\"ahler variety $X$. Let $T_0=1,T_1, \cdots, T_m$ be the basis of the cohomology group $H^*(X,\bZ)$, and $T_1, \cdots, T_r$ be the basis of the second cohomology group $H^2(X,\bZ)$. We define the matrix $g_{ij}=\int_{X}T_i\cup T_j$, and its inverse matrix $g^{ij}=(g_{ij})^{-1}$, which provide the dual basis 
\bea
T^a=\sum_{b=1}^mg^{ab}T_b~,
\eea
so that $\int_X T^i\cup T_j=\delta^i_j$. We denote by $\overline\cM_{g,n}(X,\beta)$ the moduli space of stable maps from connected genus
$g$ curves with $n$-marked points to $X$ representing the class $\beta\in H_2(X,\bZ)$. Let $\cL_1,\cdots,\cL_n$ be the corresponding tautological line bundles over $\overline\cM_{g,n}(X,\beta)$. For $\gamma_1,\cdots,\gamma_n\in H^*(X,\bZ)$ and non-negative integers $d_i$, the gravitational correlation function is defined
\bea\label{grav-correl}
\left\langle \tau_{d_1} \gamma_1, \cdots,  \tau_{d_n} \gamma_n \right\rangle_{g,\beta}=\int_{[\overline\cM_{g,n}(X,\beta)]^{\textrm{vir}}} \prod_{i=1}^n c_1(\cL_i) ^{d_i}\cup\textrm{ev}^*(\gamma_i)~.
\eea
The $J$-function of $X$ is defined by using the \emph{psi class} $\psi=c_1(\cL_1)$
\bea
J(X)=e^{\delta/\hbar}\left(1+\sum_{\beta\in H_2(X,\bZ)}\sum_{a=1}^m q^\beta \left\langle\frac{T_a}{\hbar-\psi},1 \right\rangle_{0,\beta}~T^a\right)~,
\eea
where  $\delta=\sum_{i=1}^r t_i T_i$ and $q^\beta=e^{\int_\beta \delta}$. Thus, it is regarded as a generating function for once-punctured genus zero Gromov-Witten invariants with gravitational descendants.

Now, let us compute the partition function of the 2d gauge theory on $S^2$. The Coulomb branch formula of the partition function is given by
\begin{eqnarray}
Z[\Fl_N]&=&\dfrac{1}{1!\cdots (N-1)!} \sum_{\substack{\vec{B}^{(I)} \\ I=1\cdots N-1} }\int \prod_{I=1}^{N-1}\prod_{s=1}^{I} \frac{d\tau_{s}^{(I)}}{2\pi i}e^{4\pi \xi^{(I)} \tau_{s}^{(I)} - i \theta^{(I)} B_s^{(I)}} Z_{\text{vector}} Z_{\text{bifund}} Z_{\text{fund}} ~,\cr
Z_{\text{vector}} &=& \prod_{I=2}^{N-1}\prod_{s<t}^{I} \left( \tfrac{(B_{st}^{(I)})^2}{4} - (\tau_{st}^{(I)})^2  \right)~,\cr
Z_{\text{bifund}} &=& \prod_{I=1}^{N-2} \prod_{s=1}^{I}\prod_{t=1}^{{I+1}} \tfrac{\Gamma\left(\tau_{s}^{(I)} - \tau_{t}^{(I+1)} -\tfrac{B_s^{(I)}}{2} + \tfrac{B_t^{(I+1)}}{2}\right)}{\Gamma\left(1-\tau_{s}^{(I)} + \tau_{t}^{(I+1)} -\tfrac{B_s^{(I)}}{2} + \tfrac{B_t^{(I+1)}}{2}\right)}~, \cr
Z_{\text{fund}} &=& \prod_{s=1}^{N-1}\prod_{t=1}^{N} \tfrac{\Gamma\left(\tau_{s}^{(N-1)} -\tfrac{B_s^{(N-1)}}{2}-\hbar^{-1}a_t\right)}{\Gamma\left(1-\tau_{s}^{(N-1)} -\tfrac{B_s^{(N-1)}}{2}+\hbar^{-1}a_t\right)}~,
\end{eqnarray}  
where $\xi^{(I)}$ is the Fayet-Iliopoulos parameter, $\theta^{(I)}$ is the theta angle and  $B_s^{(I)}$ are quantized magnetic fluxes on $S^2$ associated to the gauge group $\U(I)$. In the integrand, the gamma functions have an infinite tower of poles at negative integers. These towers of poles can be dealt by making changes of variables
\begin{equation}
\tau_{s}^{(I)} = \frac{B_s^{(I)}}{2} - \ell_s^{(I)}+\hbar^{-1}a_s - \hbar^{-1}H_{s}^{(I)}~,
\end{equation}
where $\ell_s^{(I)}$ are non-negative integers. Defining $k_s^{(I)}=\ell_s^{(I)}-B_s^{(I)}$, the summation can be written as $\sum_{\vec{B}^{(I)}\in \bZ}\sum_{\vec{\ell}^{(I)}\ge 0}=\sum_{\vec{k}^{(I)}\ge 0}\sum_{\vec{\ell}^{(I)}\ge 0}$ so that
one can manipulate the partition function into
\begin{eqnarray}\label{fl-contour2}
Z[\Fl_N]&=&\dfrac{1}{1!\cdots (N-1)!}\cr
&&  \sum_{\sigma\in S_N} \oint \prod_{I=1}^{N-1}\prod_{s=1}^{I}  \frac{-d  H_{s}^{(I)}}{2\pi \hbar i} (z_I\overline z_I)^{ \hbar^{-1} \vert H^{(I)} \vert-\hbar^{-1}\sum_{t=1}^I a_{\sigma(t)}} \wt Z_{\text{1-loop}} (a_{\sigma(i)})\wt Z_{\text{v}} (a_{\sigma(i)}) \wt Z_{\text{av}}(a_{\sigma(i)})~,\cr
\wt Z_{\text{1-loop}} &=&\hbar^{2\hbar^{-1} \left[ \sum_{I=1}^{N-2} (\vert H^{(I+1)} \vert {I} -\vert H^{(I)} \vert ({I+1}) ) - N \vert H^{(N-1)} \vert \right] } \prod_{I=2}^{N-1}\prod_{s\neq t}^{I}  \gamma\left(1-\hbar^{-1}H_{st}^{(I)} +\hbar^{-1}a_{st}\right)\cr
&&\prod_{I=1}^{N-2} \prod_{s=1}^{I}\prod_{t=1}^{{I+1}} \gamma\left(\hbar^{-1} H_{t}^{(I+1)}-\hbar^{-1} H_{s}^{(I)}  +\hbar^{-1}a_{st}\right)  \prod_{s=1}^{N-1}\prod_{t=1}^{N} \gamma\left(-\hbar^{-1} H_{s}^{(N-1)}+\hbar^{-1}a_{st}\right)~,\cr
\wt Z_{\text{v}} &=& \sum_{\vec{k}^{(I)}\ge0} \hbar^{- \sum_{I=1}^{N-1} \vert k^{(I)} \vert  } \prod_{I=1}^{N-1}z_I^{\vert k^{(I)}\vert }  \prod_{I=2}^{N-1}\prod_{s\neq t}^{I} \tfrac{1}{(\hbar^{-1} H_{st}^{(I)}-\hbar^{-1}a_{st})_{k_s^{(I)}-k_t^{(I)}}} \cr
&& \prod_{I=1}^{N-2} \prod_{s=1}^{I}\prod_{t=1}^{{I+1}} \tfrac{1}{(1+\hbar^{-1} H_{s}^{(I)} -\hbar^{-1} H_{t}^{(I+1)}-\hbar^{-1}a_{st})_{k^{(I)}_s - k^{(I+1)}_t}}\prod_{s=1}^{N-1}\prod_{t=1}^{N}\tfrac{1}{(1+\hbar^{-1} H_{s}^{(N-1)}-\hbar^{-1}a_{st})_{k^{(N-1)}_s}} ~,\cr
\wt Z_{\text{av}} &=& \sum_{\vec{\ell}^{(I)}\ge0} (-\hbar)^{- \sum_{I=1}^{N-1} \vert \ell^{(I)} \vert  \vert } \prod_{I=1}^{N-1}\overline z_I^{\vert \ell^{(I)}\vert }  \prod_{I=2}^{N-1}\prod_{s\neq t}^{I} \tfrac{1}{(\hbar^{-1} H_{st}^{(I)}-\hbar^{-1}a_{st})_{\ell_s^{(I)}-\ell_t^{(I)}}}\cr
&& \prod_{I=1}^{N-2} \prod_{s=1}^{I}\prod_{t=1}^{{I+1}} \tfrac{1}{(1+\hbar^{-1} H_{s}^{(I)} -\hbar^{-1} H_{t}^{(I+1)}-\hbar^{-1}a_{st})_{\ell^{(I)}_s - \ell^{(I+1)}_t}} \prod_{s=1}^{N-1}\prod_{t=1}^{N}\tfrac{1}{(1+\hbar^{-1} H_{s}^{(N-1)}-\hbar^{-1}a_{st})_{\ell^{(N-1)}_s}}~,\cr
&&
\eea
where $z_I=e^{-2\pi \xi^{(I)} +i\theta^{(I)}}$. (See \cite{Bonelli:2013mma} for more detail.)
In addition,  here we define
\bea
\g(x):=\frac{\Gamma(x)}{\Gamma(1-x)}~,
\eea
and the Pochhammer symbol $(x)_k$ is defined as
\begin{equation}
(x)_k = \left\{ 
\begin{array}{cl}
\prod_{i=0}^{k-1} (x+i) & \,\,\text{for}\,\, k>0\\
1 & \,\,\text{for}\,\, k=0\\
\prod_{i=1}^{k} \dfrac{1}{x-i} & \,\,\text{for}\,\, k<0~.
\end{array}
\right. \label{poc}
\end{equation}
As shown in \cite{Bonelli:2013mma}, the vortex partition function in the massless limit $a_s=0$ is identical with the Givental $J$-function of the complete flag variety \cite{Bertram:2004}
\bea\label{J-fl}
J[\Fl_N]&=&\sum_{\vec{k}^{(I)}} \hbar^{- \sum_{I=1}^{N-1} \vert k^{(I)} \vert   } \prod_{I=1}^{N-1}z_I^{\vert k^{(I)}\vert }  \prod_{I=2}^{N-1}\prod_{s\neq t}^{I} \tfrac{1}{(\hbar^{-1} H_{st}^{(I)})_{k_s^{(I)}-k_t^{(I)}}} \\
&& \prod_{I=1}^{N-2} \prod_{s=1}^{I}\prod_{t=1}^{{I+1}} \tfrac{1}{(1+\hbar^{-1} H_{s}^{(I)} -\hbar^{-1} H_{t}^{(I+1)})_{k^{(I)}_s - k^{(I+1)}_t}}\prod_{s=1}^{N-1}\prod_{t=1}^{N}\tfrac{1}{(1+\hbar^{-1} H_{s}^{(N-1)}-\hbar^{-1} H_{t}^{(N)})_{k^{(N-1)}_s}} ~.\nonumber
\eea
Here we identify $H_{s}^{(I)}$ $(s = 1,...,I)$ with Chern roots to the duals of the
universal bundles $\cS_I$:
\be
0\subset \cS_1 \subset \cS_2 \subset \dots \subset \cS_{N-1} \subset \cS_{N} = 
{\bC}^N \otimes {\cal O}_{\Fl_N}~.
\ee 
and we add $H_{t}^{(N)}$ $(t=1,\cdots,N)$ to the last Pochhammer of $\wt Z_v$ by hand. These additional classes are necessary to become an eigenfunction of the Toda Hamiltonian as we will see below.

Performing the residue integral in \eqref{fl-contour2}, one obtains the Higgs branch formula
\bea
Z[\Fl_N]&=&\dfrac{1}{1!\cdots (N-1)!}  \sum_{\sigma\in S_N}  \prod_{I=1}^{N-1} (z_I\overline z_I)^{-\hbar^{-1}\sum_{t=1}^I a_{\sigma(t)}}  Z_{\text{1-loop}} (a_{\sigma(i)}) Z_{\text{v}} (a_{\sigma(i)})  Z_{\text{av}}(a_{\sigma(i)})~,\cr
 Z_{\text{1-loop}} &=&\prod_{s< t}^N \gamma\left(\frac{a_s-a_t}{\hbar}\right)~,\cr
 Z_{\text{v}} &=& \sum_{\vec{k}^{(I)}} \hbar^{- \sum_{I=1}^{N-1} \vert k^{(I)} \vert } \prod_{I=1}^{N-1}z_I^{\vert k^{(I)}\vert } \prod_{I=2}^{N-1}\prod_{s\neq t}^{I} \tfrac{1}{(-\hbar^{-1}a_{st})_{k_s^{(I)}-k_t^{(I)}}}  \cr
&&\prod_{I=1}^{N-2} \prod_{s=1}^{I}\prod_{t=1}^{{I+1}} \tfrac{1}{(1-\hbar^{-1}a_{st})_{k^{(I)}_s - k^{(I+1)}_t}}\prod_{s=1}^{N-1}\prod_{t=1}^{N}\tfrac{1}{(1-\hbar^{-1}a_{st})_{k^{(N-1)}_s}} ~,\cr
 Z_{\text{av}} &=& \sum_{\vec{\ell}^{(I)}} (-\hbar)^{- \sum_{I=1}^{N-1} \vert \ell^{(I)} \vert  }\prod_{I=1}^{N-1}\overline z_I^{\vert \ell^{(I)}\vert }  \prod_{I=2}^{N-1}\prod_{s\neq t}^{I} \tfrac{1}{(-\hbar^{-1}a_{st})_{\ell_s^{(I)}-\ell_t^{(I)}}}\cr
 &&  \prod_{I=1}^{N-2} \prod_{s=1}^{I}\prod_{t=1}^{{I+1}} \tfrac{1}{(1-\hbar^{-1}a_{st})_{\ell^{(I)}_s - \ell^{(I+1)}_t}} \prod_{s=1}^{N-1}\prod_{t=1}^{N}\tfrac{1}{(1-\hbar^{-1}a_{st})_{\ell^{(N-1)}_s}}~.
\eea
It turns out that the vortex partition function can be obtained from the instanton partition function by setting the instanton number $k=k_N=0$ 
\bea\label{flag-id}
 Z_{\text{v}}[\Fl_N] (z_I,a,\hbar)=\sum_{k_1,\cdots,k_{N-1}}  \Big( \prod_{I=1}^{N-1} z_I^{k_I} \Big) \scZ_{[1^N],k_1,\cdots,k_{N-1}, k_N=0}^{\textrm{pure}}(a,\e_1=\hbar)
\eea
where $\scZ_{[1^N],k_1,\cdots,k_{N-1}, k_N=0}^{\textrm{pure}}$ is independent of $\e_2$. This implies that the 4d instanton partition function receives the contribution only from 2d dynamics when $k=0$. In other words, the vortex partition function can be regarded as the generating function of the equivariant cohomology of the Laumon space \cite{Feigin:2008} which can be described by the hand-saw quiver \cite{Finkelberg:2010}. Moreover, the left hand side of \eqref{flag-id} has been computed from the $\cN=(2,2)$  GLSM description of the 2d theory coupled to the pure Yang-Mills whereas the description of the surface operator by the boundary condition of the gauge field has led to the right hand side. Thus, the identity \eqref{flag-id} proves that the two descriptions for the surface operator are equivalent \cite{Gukov:2006jk}.

It is straightforward to see from \eqref{periodic-toda} that the vortex partition function becomes an eigenfunction of the Toda Hamiltonian
\bea
\Big(\hbar^2\Delta_{\frakh}-2\sum_{\alpha \in \Pi}  e^{\langle t,\alpha\rangle}\Big) \Big[e^{-\frac{\langle a, t\rangle}{\hbar}}  Z_{\text{v}}[\Fl_N]  \Big]= \langle a,a\rangle\Big[e^{-\frac{\langle a, t\rangle}{\hbar}}  Z_{\text{v}}[\Fl_N]  \Big]~,
\eea
where we substitute $z_I=e^{t_{I}-t_{I+1}}$.
In addition, it is well-known that the $J$-function of the complete flag variety becomes an eigenfunction of the Toda Hamiltonian \cite{Kim:1999}. To see that, one has to identify $H^{(I)}_s=H^{(I+1)}_s$ $(s=1,\cdots,I)$ as the same cohomology class. Then, the $J$-function becomes equivalent to the generating function $ Z_{\text{v}}[\Fl_N] $ of the equivariant cohomology of the Laumon space by setting $H_s=-a_s$.

\subsubsection{One-loop determinant}\label{sec:1-loop-pure}
The localization technique enables us to demonstrate exact evaluations of supersymmetric partition functions by taking only the quadratic fluctuations over BPS configurations into account. In the case of $\cN=2$ supersymmetric gauge theories on $S^4_b$, the BPS configurations correspond to the instantons at the north and south pole of $S^4_b$ \cite{Pestun:2007rz,Hama:2012bg}. Then, the quadratic fluctuations over the instanton configurations can be evaluated by the means of the Atiyah-Singer index theory for transversally elliptic operators. The minimum explanation is provided in Appendix \ref{sec:1-loop}.

Since the field content of the $\cN=2$ pure Yang-Mills consists only of the vector multiplet, the quadratic fluctuations of the theory is captured just by the one-loop determinant \eqref{vector-1l} of the vector multiplet over the instanton configurations, which can be obtained by the equivariant indices of the self-dual \eqref{eq:self-dual} and anti-self-dual complex 
\bea
\scZ_{\textrm{1-loop}}^{\textrm{pure}}&=&\prod_{\a\in\Delta} \left[\Gamma_2\left(\langle a, \alpha \rangle|\e_1,\e_2\right)\Gamma_2\left(\langle a, \alpha \rangle+\e_1+\e_2|\e_1,\e_2\right)\right]^{-1}~,\cr
&=&\prod_{\a\in\Delta}\Upsilon\left(\langle a, \alpha \rangle|\e_1,\e_2\right)~,
\eea
where $\Delta$ represents the set of roots of $\fraksl(N)$. Note that $\Gamma_2(x|\e_1,\e_2)$ is the Barnes double Gamma function \eqref{Barnes} and $\Upsilon(x|\e_1,\e_2)$ is the Upsilon function \eqref{Upsilon}.

Since the instanton partition function in the presence of a surface operator have been computed by the orbifold operation, it is natural to expect that the one-loop computation can be obtained by the index theorem on  $\bC\times (\bC/\bZ_N)$. As in the case of the instanton partition function, the equivariant parameters are shifted by 
\be\label{equiv shift 2}
\e_2\to\frac{\e_2}{N}~,\qquad a_i\to a_i-\frac{i-1}N\e_2~,
\ee
due to the orbifold operation. This re-parametrization alters the one-loop determinant
\bea
&&\prod_{\a\in\Delta} \Gamma_2\left(\langle a, \alpha \rangle|\e_1,\e_2\right)\Gamma_2\left(\langle a, \alpha \rangle+\e_1+\e_2|\e_1,\e_2\right)\cr
&\to&\prod_{i,j=1,i\neq j}^N \G_2\left(a_i-a_j +\tfrac{j-i}{N}\e_2|\e_1,\tfrac{\e_2}N\right)\G_2\left(a_i-a_j+\e_1+\tfrac{1+j-i}{N}\e_2 |\e_1,\tfrac{\e_2}N\right)~.
\eea
To get its $\bZ_N$-invariant part, we average over the finite group $\bZ_N$ as in \eqref{average}, leaving the one-loop determinant in the existence of the full surface operator
\bea\label{vector HR}
\scZ_{\textrm{1-loop}}^{\textrm{pure}}[1^N]&=&\prod_{i,j=1,i\neq j}^N \left[\G_2(a_i-a_j+\left\lceil\tfrac{j-i}{N}\right\rceil\e_2 |\e_1,\e_2) \G_2(a_i-a_j +\e_1+\left\lceil\tfrac{1+j-i}{N}\right\rceil\e_2 |\e_1,\e_2)\right]^{-1}\cr
&=&\prod_{i,j=1,i\neq j}^N\Upsilon\left(a_i-a_j+\left\lceil\tfrac{j-i}{N}\right\rceil\e_2 |\e_1,\e_2\right)~,
\eea
where $\lceil x\rceil$ denotes the smallest integer  $\ge x$.

As we have seen in the previous sections, the instanton partition function contains both 4d and 2d dynamics, and the 2d vortex partition function is left when the 4d non-perturbative effect is switched off. This should be true for the perturbative contributions. Namely, if the 4d contribution $\scZ_{\textrm{1-loop}}^{\textrm{pure}}$ is subtracted from the one-loop determinant $\scZ_{\textrm{1-loop}}^{\textrm{pure}}[1^N]$, only the 2d effect $Z_{\textrm{1-loop}}[\Fl_N]$ should be evident \cite[\S6]{Doroud:2012xw}. In fact, using the shift relation \eqref{eq:updif} of the Upsilon function, one can see that the ratio of $\scZ_{\textrm{1-loop}}^{\textrm{pure}}[1^N]$ to $\scZ_{\textrm{1-loop}}$ is independent of $\e_2$, and we have 
\bea\label{1-loop-quotient-pure}
\frac{\scZ_{\textrm{1-loop}}^{\textrm{pure}}[1^N]}{\scZ_{\textrm{1-loop}}^{\textrm{pure}}}(a,\e_1=\hbar)=\prod_{\alpha \in \Delta^+} \hbar^{\frac{\langle a, \alpha \rangle}{\hbar}-1}\gamma\left(\frac{\langle a, \alpha \rangle}{\hbar}\right) ``="Z_{\textrm{1-loop}}[\Fl_N](a,\hbar)
\eea
where $``="$ means the equality up to a constant.\footnote{The author would like to thank Hee-Cheol Kim for suggesting this approach.} This supports the validity of the orbifold method even in the one-loop computations.

\subsection{$\cN=2^\ast$ theory}
\subsubsection{Instanton partition function}
The $\cN=2^\ast$ theory is the deformation of the $\cN=4$ SCFT by adding the mass $\mu_{\textrm{adj}}$ to the hypermultiplet in the adjoint representation. From the 6d perspective, the $\SU(N)$ $\cN=2^\ast$ theory is obtained by wrapping $N$ M5-branes on a once-punctured torus. Because the standard ADHM description of the $\cN=2^\ast$ theory \cite{Bruzzo:2002xf} can be generalized to the orbifold space $\bC\times (\bC/\bZ_N)$, one can write the contour integral representation of the $\U(N)$ instanton partition function of the $\cN=2^\ast$ theory with a full surface operator:
\bea
\scZ_{\textrm{inst}}^{\cN=2^\ast }[1^N]=\sum_{\vec{k}} \Big( \prod_{I=1}^N z_I^{k_I} \Big) \scZ_{[1^N],\vec{k}}^{\cN=2^\ast }~,
\eea
where
\bea
\scZ_{[1^N],\vec{k}}^{\cN=2^\ast } &=&\left[\frac{\e_1-\mu_{\textrm{adj}}}{\e_1 \mu_{\textrm{adj}}}\right]^{\sum_{I=1}^N k_I}   \cr
&&\oint  \prod_{I=1}^N \prod_{s=1}^{k_I}d \phi_s^{(I)} \frac{(\phi_s^{(I)} + a_I-\frac{(I-1)\e_2}{N}+\mu_{\textrm{adj}}) (\phi_s^{(I)} +a_{I+1}+\e-\frac{I\e_2}{N}-\mu_{\textrm{adj}})}{ (\phi_s^{(I)} + a_I-\frac{(I-1)\e_2}{N}) (\phi_s^{(I)} +a_{I+1}+\e-\frac{I\e_2}{N})} \cr
&& \prod_{I=1}^N \prod_{s=1}^{k_I} \prod_{t \neq s}^{k_I} \frac{ \phi_{st}^{(I)} (\phi_{st}^{(I)} + \epsilon_1-\mu_{\textrm{adj}})}{(\phi_{st}^{(I)}+\mu_{\textrm{adj}} )(\phi_{st}^{(I)} + \epsilon_1)} \cr
&&\prod_{I=1}^N \prod_{s=1}^{k_I} \prod_{t=1}^{k_{I+1}} \frac{(\phi_{s}^{(I)} - \phi_t^{(I+1)} + \epsilon)(\phi_{s}^{(I)} - \phi_t^{(I+1)} + \frac{\epsilon_2}{N}-\mu_{\textrm{adj}})}{(\phi_{s}^{(I)} - \phi_t^{(I+1)} + \frac{\epsilon_2}{N})(\phi_{s}^{(I)} - \phi_t^{(I+1)} + \epsilon-\mu_{\textrm{adj}})}~.
\eea
It was proven in \cite{Negut:2011} that, by multiplying an appropriate factor, the instanton partition function $\scZ_{\textrm{inst}}^{\cN=2^\ast }[1^N]$ becomes an eigenfunction of a non-stationary deformation of the trigonometric Calogero-Moser Hamiltonian. To avoid repetition, we refer the reader to \cite{Negut:2011} for the explicit expression of the differential equation. Instead, let us mention the connection to the Knizhnik-Zamolodchikov-Bernard (KZB) equation \cite{Knizhnik:1984nr,Bernard:1987df,Felder:1995}. 

In the AGT relation, the partition function of the $\cN=2^*$ theory is dual to the one-point correlation function on a torus. When a full surface operator is present, the instanton partition function of  the $\cN=2^*$ theory is the one-point $\wh \fraksl(N)$ conformal blocks on a torus. More precisely, the corresponding conformal block is a semi-degenerate field $\bV_{\kappa\omega_{N-1}}(x;q)$ on a torus with the $\cK$ operator \cite{Negut:2011,Alday:2010vg,Kozcaz:2010yp}
\bea
F_{\cK}(x;q):=\Tr_{\bV_j} \cK(x;q) \bV_{\kappa\omega_{N-1}}(x;q)~,
\eea
where $\bV_j$ is the Verma module of the affine Lie algebra $\wh\fraksl(N)$ with the highest weight $j$. Note that the  semi-degenerate field $\bV_{\kappa\omega_{N-1}}(x;q)$ labelled by the momentum proportional to the fundamental weight $\omega_{N-1}$ depends on the  isospin variables $x_i$ $(i=1,\cdots,N-1)$ and the world-sheet variable $q$. We refer the reader to \cite{Alday:2010vg,Kozcaz:2010yp} for the explicit expression of  the $\cK$ operator. Writing the instanton partition function in terms of $q=e^{2\pi i \tau}$ and $t_i$ $(i=1,\cdots,N-1)$ via \eqref{cov}, it is conjectured that it matches with the $\wh \fraksl(N)$ conformal block up to the $\U(1)$ factor
\bea
\prod_{i=1}^\infty (1-q^{i})^{-\frac{\mu_{\textrm{adj}}(N\e_1+\e_2-N\mu_{\textrm{adj}})}{\e_1\e_2}+1}\scZ_{\textrm{inst}}^{\cN=2^\ast }[1^N]=F_{\cK}(x_\ell=e^{t_1-t_{\ell+1}};q)
\eea
Here, the parameters are identified by
\bea\label{torus-id}
  \frac{a}{\e_1}=j+\rho~, \qquad \frac{\mu_{\textrm{adj}}}{\e_1}=-\frac{\kappa}{N}~,\qquad -\frac{\e_2}{\e_1}=k+N~,
\eea
where $\rho$ is the Weyl vector and $k$ is the level. We further conjecture that, for the once-punctured conformal block on a torus,  the effect of the insertion of the $\cK$ operator results in the prefactor so that the ordinary conformal block  $F(x;q):=\Tr_{\bV_j} \bV_{\kappa\omega_{N-1}}(x;q)$ is proportional to $F_{\cK}(x;q)$ 
\bea\label{prefactor}
F(x_\ell=e^{t_1-t_{\ell+1}};q)=  f(t,q)^{\frac{\kappa}{N}}  F_{\cK}(x_\ell=e^{t_1-t_{\ell+1}};q)~.
\eea
When $N=2$, the explicit expression of the prefactor is found by computer analysis, which is $f(t,q)=1-e^{t_1-t_2}-q e^{t_2-t_1}$ \cite[(4.20)]{Alday:2010vg}. We expect that this relation holds for higher rank gauge groups.
Then, taking into account this prefactor and the $\U(1)$ factor, we can define the function
\be\label{calY}
\scY(t,q,a,\mu_{\textrm{adj}},\e_1,\e_2):=e^{-\frac{\langle a,t\rangle}{\e_1}}f(t,q)^{-\frac{\mu_{\text{adj}}}{\e_1}+1}\prod_{i=1}^\infty (1-q^{i})^{-\frac{\mu_{\textrm{adj}}(N\e_1+\e_2-N\mu_{\textrm{adj}})}{\e_1\e_2}+1}\scZ_{\textrm{inst}}^{\cN=2^\ast }[1^N]
\ee
so that it should satisfy the KZB equation
\be\label{KZB}
\left[2\e_1\e_2q \frac{\p}{\p q}+\e_1^2\Delta_{\frakh} + 2\mu_{\textrm{adj}}(\mu_{\textrm{adj}}-\e_1) \sum_{\a\in\Delta^+}\Big(\frac{1}{4\pi^2}\wp({\langle t,\alpha\rangle};\t)+\frac{1}{12}\Big)\right]\scY=\langle a,a\rangle \scY~.
\ee
Note that  the Weierstrass elliptic function $\wp(u;\t)$ can be expressed as \cite[\S 8.5]{Dabholkar:2012nd}
\be \label{pe}
\frac{1}{4\pi^{2}} \wp(u;\t)   \=  T_{2}(u;\t) - \frac{1}{12}E_{2}(\t) \, ,
\ee
where $E_{2}(\t)$ is the Eisenstein series 
\be
E_2(\t)=1-\frac{2}{3}\sum_{n= 1}^\infty \frac{nq^n}{1-q^n}~,
\ee
and we define
\bea \label{T2}
T_{2}(u;\t) := - \sum_{\ell \in \bZ}  \frac{q^{\ell} \, e^{u}}{(1 - q^{\ell} \, e^{ u})^{2}} ~.
\eea
Since the solution \eqref{calY} of the KZB equation should reduce to the eigenfunction \eqref{Y} of the trigonometric Calogero-Moser Hamiltonian at $q= 0$, the $q= 0$ specialization of the prefactor is 
\bea
f(t,q=0)= \prod_{\alpha\in \Delta^+} (1-e^{\langle t,\alpha\rangle})~.
\eea
When $N=2$, the prefactor is subject to this condition. Nevertheless, it is crucial to find the explicit expression of the prefactor $f(t,q)$ for a higher rank gauge group. Moreover, due to the insertion of the  $\cK$ operator, it is not obvious that the instanton partition functions of class $\cS$ theories generally satisfy the KZ equations \cite{Knizhnik:1984nr}. Therefore, it is valuable to gain a better understanding of the meaning of the $\cK$ operator.

\subsubsection{$J$-function of cotangent bundle of complete flag variety}\label{sec:cotangent}

\begin{figure}[h]
 \centering
 \includegraphics[width=10cm]{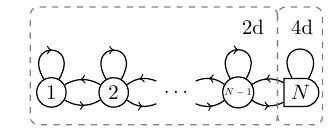}
    \caption{Quiver diagram of the 2d-4d coupled system for the $\cN=2^*$ theory in the presence of a full surface operator. The Higgs branch of the $\cN=(2,2)^*$ GLSM  is the cotangent bundle of the complete flag variety.}\label{fig:cotangent}
\end{figure}

Since the $\cN=2^*$ theory is a mass deformation of the $\cN=4$ SCFT, the dynamics on the support of surface operator is also described by a deformation of an $\cN=(4,4)$ supersymmetric gauge theory specified by the quiver diagram  (Figure \ref{fig:cotangent}) where the matter content is given as follows:
\begin{itemize}
\item bifundamentals $Q^{(I)} \in  ( \boldsymbol I,\boldsymbol{\overline{I+1}})$, $\wt Q^{(I)} \in  (\boldsymbol{ \overline{I}},\boldsymbol{{I+1}})$ $\hspace{1cm}$ $(I\in 1,\cdots, N-2)$
\item one adjoint $\Phi^{(I)}$ for each gauge group $\U(I)$ $\hspace{1cm}$ $(I\in 1,\cdots, N-1)$
\item $N$ fundamentals $Q^{(N-1)}$ and $N$ antifundamentals $\wt Q^{(N-1)}$  of $\U(N-1)$
\end{itemize}
The theory is the deformation of the $\cN=(4,4)$ supersymmetric gauge theory with the superpotential
\bea
W= \sum_{I=1}^{N-1} \Tr\; \wt Q^{(I)}\Phi^{(I)} Q^{(I)}+ \sum_{I=1}^{N-2}\Tr\;  Q^{(I)}\Phi^{(I+1)} \wt Q^{(I)}~,
\eea
by turning on the twisted mass $m$ of $\wt Q^{(I)}$ and $\Phi^{(I)}$ ($I=1,\cdots,N-1)$. Note that the $R$-charges of  $\Phi^{(I)}$ are two and those of   $Q^{(I)}$ and $\wt Q^{(I)}$ are zero.
The infrared dynamics of this theory is described by the hyper-K\"ahler NLSM with the contangent bundle $T^*\Fl_N$ of the complete flag variety \cite{Gukov:2006jk}. Let us first compute the exact partition function of this theory without turning on the twisted masses coming from the Coulomb branch parameters. The Coulomb branch formula of the partition function is given by
\begin{eqnarray}\label{cotangent-flag}
Z[T^*\Fl_N]&=&\dfrac{1}{1!\cdots (N-1)!} \cr
&&\times\sum_{\substack{\vec{B}^{(I)} \\ I=1\cdots N-1} }\int \prod_{I=1}^{N-1}\prod_{s=1}^{I} \frac{d\tau_{s}^{(I)}}{2\pi i}e^{4\pi \xi^{(I)} \tau_{s}^{(I)} - i \theta^{(I)} B_s^{(I)}} Z_{\text{vect}} Z_{\text{adj}}Z_{\text{bifund}} Z_{\text{fund-anti}} ~,\cr
Z_{\text{vect}} &=& \prod_{I=2}^{N-1}\prod_{s<t}^{I} \left( \frac{(B_{st}^{(I)})^2}{4} - (\tau_{st}^{(I)})^2  \right) ~,\cr
Z_{\text{bifund}} &=& \prod_{I=1}^{N-2} \prod_{s=1}^{I}\prod_{t=1}^{{I+1}} \tfrac{\Gamma\left(\tau_{s}^{(I)} - \tau_{t}^{(I+1)} -\tfrac{B_s^{(I)}}{2} + \tfrac{B_t^{(I+1)}}{2}\right)\Gamma\left(-\tau_{s}^{(I)} + \tau_{t}^{(I+1)} +\tfrac{B_s^{(I)}}{2} - \tfrac{B_t^{(I+1)}}{2}-\hbar^{-1}m\right)}{\Gamma\left(1-\tau_{s}^{(I)} + \tau_{t}^{(I+1)} -\tfrac{B_s^{(I)}}{2} + \tfrac{B_t^{(I+1)}}{2}\right)\Gamma\left(1+\tau_{s}^{(I)} - \tau_{t}^{(I+1)} +\tfrac{B_s^{(I)}}{2} - \tfrac{B_t^{(I+1)}}{2}+\hbar^{-1}m\right)} ~,\cr
Z_{\text{fund-anti}} &=& \prod_{s=1}^{N-1}\left[ \tfrac{\Gamma\left(\tau_{s}^{(N-1)} -\tfrac{B_s^{(N-1)}}{2}\right)\Gamma\left(-\tau_{s}^{(N-1)} +\tfrac{B_s^{(N-1)}}{2}-\hbar^{-1}m\right)}{\Gamma\left(1-\tau_{s}^{(N-1)} -\tfrac{B_s^{(N-1)}}{2}\right)\Gamma\left(1+\tau_{s}^{(N-1)} +\tfrac{B_s^{(N-1)}}{2}+\hbar^{-1}m\right)}\right]^{N}~,\cr
Z_{\text{adj}} &=&\prod_{I=2}^{N-1}\prod_{s\neq t}^{I}  \tfrac{\Gamma\left(1+\tau_{st}^{(I)} -\tfrac{B_{st}^{(I)}}{2}+\hbar^{-1}m\right)}{\Gamma\left(-\tau_{st}^{(I)} -\tfrac{B_{st}^{(I)}}{2}-\hbar^{-1}m\right)}~.
\end{eqnarray}  
Defining 
\begin{equation}
\tau_{s}^{(I)} = \frac{B_s^{(I)}}{2} - \ell_s^{(I)} -\hbar^{-1}H_{s}^{(I)}~,
\end{equation}
the same manipulation as in \eqref{fl-contour2} yields
\begin{eqnarray}
Z[T^*\Fl_N]&=&\dfrac{1}{1!\cdots (N-1)!} \oint \prod_{I=1}^{N-1}\prod_{s=1}^{I}  \frac{-d  H_{s}^{(I)}}{2\pi \hbar i} (z_I\overline z_I)^{ \hbar^{-1} \vert H^{(I)} \vert}  \wt Z_{\text{1-loop}}\wt Z_{\text{v}}\wt Z_{\text{av}}\\
\wt Z_{\text{1-loop}} &=& \prod_{I=2}^{N-1}\prod_{s\neq t}^{I}  \gamma\left(1-\hbar^{-1}H_{st}^{(I)}\right)\gamma\left(1+\hbar^{-1}H_{st}^{(I)}+\hbar^{-1}m \right)~,\cr
&&\prod_{I=1}^{N-2} \prod_{s=1}^{I}\prod_{t=1}^{{I+1}} \gamma\left(-\hbar^{-1} H_{s}^{(I)} +\hbar^{-1} H_{t}^{(I+1)} \right)\gamma\left(\hbar^{-1} H_{s}^{(I)} -\hbar^{-1} H_{t}^{(I+1)}-\hbar^{-1}m \right) ~,\cr
&& \prod_{s=1}^{N-1}\left[ \gamma\left(-\hbar^{-1} H_{s}^{(N-1)}\right)\gamma\left(\hbar^{-1} H_{s}^{(N-1)}-\hbar^{-1}m\right)\right]^{N} ~,\cr
\wt Z_{\text{v}} &=& \sum_{\vec{k}^{(I)}}  \prod_{I=1}^{N-1}z_I^{\vert k^{(I)}\vert } \prod_{I=2}^{N-1}\prod_{s\neq t}^{I} \tfrac{(1+\hbar^{-1} H_{st}^{(I)}+\hbar^{-1}m)_{k_s^{(I)}-k_t^{(I)}}}{(\hbar^{-1} H_{st}^{(I)})_{k_s^{(I)}-k_t^{(I)}}} \cr
&& \prod_{I=1}^{N-2} \prod_{s=1}^{I}\prod_{t=1}^{{I+1}} \tfrac{(\hbar^{-1} H_{s}^{(I)} -\hbar^{-1} H_{t}^{(I+1)}-\hbar^{-1}m)_{k^{(I)}_s - k^{(I+1)}_t}}{(1+\hbar^{-1} H_{s}^{(I)} -\hbar^{-1} H_{t}^{(I+1)})_{k^{(I)}_s - k^{(I+1)}_t}} \prod_{s=1}^{N-1}\left[ \tfrac{(\hbar^{-1} H_{s}^{(N-1)}-\hbar^{-1}m)_{k^{(N-1)}_s}}{(1+\hbar^{-1} H_{s}^{(N-1)})_{k^{(N-1)}_s}}\right]^{N} ~,\cr
\wt Z_{\text{av}} &=& \sum_{\vec{\ell}^{(I)}} \prod_{I=1}^{N-1}\overline z_I^{\vert \ell^{(I)}\vert}\prod_{I=2}^{N-1}\prod_{s\neq t}^{I} \tfrac{(1+\hbar^{-1} H_{st}^{(I)}+\hbar^{-1}m)_{\ell_s^{(I)}-\ell_t^{(I)}}}{(\hbar^{-1}  H_{st}^{(I)})_{\ell_s^{(I)}-\ell_t^{(I)}}} \cr
&& \prod_{I=1}^{N-2} \prod_{s=1}^{I}\prod_{t=1}^{{I+1}} \tfrac{(\hbar^{-1} H_{s}^{(I)} -\hbar^{-1} H_{t}^{(I+1)}-\hbar^{-1}m)_{\ell^{(I)}_s - \ell^{(I+1)}_t}}{(1+\hbar^{-1} H_{s}^{(I)} -\hbar^{-1} H_{t}^{(I+1)})_{\ell^{(I)}_s - \ell^{(I+1)}_t}} \prod_{s=1}^{N-1}\left[ \tfrac{(\hbar^{-1} H_{s}^{(N-1)}-\hbar^{-1}m)_{\ell^{(N-1)}_s}}{(1+\hbar^{-1} H_{s}^{(N-1)})_{\ell^{(N-1)}_s}}\right]^{N}~.\nonumber
\eea
From the expression $\wt Z_{\text{v}}$, we conjecture the $J$-function of the cotangent bundle $T^*\Fl_N$ of the complete flag variety as \footnote{After posting this paper on arXiv, Bumsig Kim informed that the formula \eqref{J-cotangent} follows as a special case of  Theorem 6.1.2 in \cite{Fontanine:2008}, 
using the quantum Lefschetz theorem \cite{Coates:2007} and \eqref{J-fl}.}
\bea\label{J-cotangent}
J[T^*\Fl_N]&=&\sum_{\vec{k}^{(I)}}  \prod_{I=1}^{N-1}z_I^{\vert k^{(I)}\vert } \prod_{I=2}^{N-1}\prod_{s\neq t}^{I} \tfrac{(1+\hbar^{-1} H_{st}^{(I)}+\hbar^{-1}m)_{k_s^{(I)}-k_t^{(I)}}}{(\hbar^{-1} H_{st}^{(I)})_{k_s^{(I)}-k_t^{(I)}}} \cr
&& \prod_{I=1}^{N-2} \prod_{s=1}^{I}\prod_{t=1}^{{I+1}} \tfrac{(\hbar^{-1} H_{s}^{(I)} -\hbar^{-1} H_{t}^{(I+1)}-\hbar^{-1}m)_{k^{(I)}_s - k^{(I+1)}_t}}{(1+\hbar^{-1} H_{s}^{(I)} -\hbar^{-1} H_{t}^{(I+1)})_{k^{(I)}_s - k^{(I+1)}_t}}\cr
&& \prod_{s=1}^{N-1}\prod_{t=1}^{N} \tfrac{(\hbar^{-1} H_{s}^{(N-1)}-\hbar^{-1} H_{t}^{(N)}-\hbar^{-1}m)_{k^{(N-1)}_s}}{(1+\hbar^{-1} H_{s}^{(N-1)}-\hbar^{-1} H_{t}^{(N)})_{k^{(N-1)}_s}}~.
\eea
It is worth mentioning that the definition given in \S \ref{sec:flag} is not appropriate for the $J$-function of the cotangent bundle of a flag variety. It appears that one has to introduce the equivariant parameter $m$ of the fiber direction somehow to its definition. The $J$-function of the cotangent bundle of a partial flag variety are given in Appendix \ref{sec:partial}.

Incorporating the twisted mass $a_s$ and performing the residue integral, one can write the Higgs branch formula
\bea
Z[T^*\Fl_N]&=&\dfrac{1}{1!\cdots (N-1)!}  \sum_{\sigma\in S_N}   \prod_{I=1}^{N-1} (z_I\overline z_I)^{-\hbar^{-1}\sum_{t=1}^I a_{\sigma(t)}}  Z_{\text{1-loop}} (a_{\sigma(i)}) Z_{\text{v}} (a_{\sigma(i)})  Z_{\text{av}}(a_{\sigma(i)})~,\cr
 Z_{\text{1-loop}} &=&\prod_{s< t}^N \gamma\left(\frac{a_s-a_t}{\hbar}\right)\gamma\left(\frac{a_t-a_s-m}{\hbar}\right)~,\cr
 Z_{\text{v}} &=& \sum_{\vec{k}^{(I)}}\prod_{I=1}^{N-1} z_I^{\vert k^{(I)}\vert } \prod_{I=2}^{N-1}\prod_{s\neq t}^{I} \tfrac{(1-\hbar^{-1} a_{st}+\hbar^{-1}m)_{k_s^{(I)}-k_t^{(I)}}}{(-\hbar^{-1} a_{st})_{k_s^{(I)}-k_t^{(I)}}} \cr
&&\prod_{I=1}^{N-2} \prod_{s=1}^{I}\prod_{t=1}^{{I+1}} \tfrac{(-\hbar^{-1}a_{st}-\hbar^{-1}m)_{k^{(I)}_s - k^{(I+1)}_t}}{(1-\hbar^{-1}a_{st})_{k^{(I)}_s - k^{(I+1)}_t}}\prod_{s=1}^{N-1}\prod_{t=1}^{N}\tfrac{(-\hbar^{-1}a_{st}-\hbar^{-1}m)_{k^{(N-1)}_s}}{(1-\hbar^{-1}a_{st})_{k^{(N-1)}_s}} ~,\cr
 Z_{\text{av}} &=& \sum_{\vec{\ell}^{(I)}}  \prod_{I=1}^{N-1} \overline z_I^{\vert \ell^{(I)}\vert } \prod_{I=2}^{N-1}\prod_{s\neq t}^{I} \tfrac{(1-\hbar^{-1} a_{st}+\hbar^{-1}m)_{\ell_s^{(I)}-\ell_t^{(I)}}}{(-\hbar^{-1} a_{st})_{\ell_s^{(I)}-\ell_t^{(I)}}} \cr
 && \prod_{I=1}^{N-2} \prod_{s=1}^{I}\prod_{t=1}^{{I+1}} \tfrac{(-\hbar^{-1}a_{st}-\hbar^{-1}m)_{\ell^{(I)}_s - \ell^{(I+1)}_t}}{(1-\hbar^{-1}a_{st})_{\ell^{(I)}_s - \ell^{(I+1)}_t}} \prod_{s=1}^{N-1}\prod_{t=1}^{N}\tfrac{(-\hbar^{-1}a_{st}-\hbar^{-1}m)_{\ell^{(N-1)}_s}}{(1-\hbar^{-1}a_{st})_{\ell^{(N-1)}_s}}~.
\eea
As in the case of the pure Yang-Mills, the vortex partition function can be obtained by setting the instanton number $k=k_N=0$ 
\bea\label{inst-zero}
&& Z_{\text{v}}[T^*\Fl_N] (z_I,a_i,m,\hbar)\\
 &&\hspace{2cm}=\sum_{k_1,\cdots,k_{N-1}}  \Big( \prod_{I=1}^{N-1} z_I^{k_I} \Big) \scZ_{[1^N],k_1,\cdots,k_{N-1}, k_N=0}^{\cN=2^*}(a_i,\mu_{\textrm{adj}}=m+\hbar,\e_1=\hbar)~.\nonumber
\eea
Multiplying the following factor to the vortex partition function
\bea\label{Y}
Y(t,a,m,\hbar)=e^{-\frac{\langle a, t\rangle}{\hbar} }  \prod_{\alpha\in \Delta^+} (1-e^{\langle t,\alpha\rangle})^{-\frac{m}{\hbar}} ~ Z_{\text{v}} [T^*\Fl_N]~,
\eea
it becomes an eigenfunction of  the trigonometric Calogero-Moser Hamiltonian \cite{Negut:2008,Braverman:2010}
\bea\label{tCM}
\left[\hbar^2\Delta_{\frakh}-2m(m+\hbar)\sum_{\alpha\in \Delta^+} \frac 1{(e^{\langle t,\alpha\rangle/2}-e^{-\langle t,\alpha\rangle/2})^2}\right]Y=\langle a,a\rangle Y~,
\eea
where $\Delta^+$ represents the set of positive roots of $\fraksl(N)$. It is easy to see from \eqref{pe} that the potential of the trigonometric Calogero-Moser Hamiltonian \eqref{tCM} can be obtained by taking $q\to 0$ limit of the potential of the elliptic Calogero-Moser Hamiltonian \eqref{KZB}. Furthermore, the monodromy matrices of this differential equation satisfy the affine Hecke algebra \cite{Braverman:2010}. This algebra admits a natural physical interpretation as the action of the loop operators on a full surface operator \cite{Gukov:2006jk,Honda:2013uca}.

\subsubsection{Twisted chiral ring}\label{sec:bethe}
Let us study the twisted chiral ring in the Landau-Ginzburg (LG) model mirror dual to the NLSM with $T^*\Fl_N$.\footnote{The twisted chiral rings of the flag varieties has been investigated in \cite{Gaiotto:2013sma} with a different approach.} When the FI parameter is negative infinity, the effective theory of the GLSM is described by the LG model. Moreover, the LG model also provides the mirror description of the NLSM. Since the detail prescription to write the $S^2$ partition function in terms of the LG model is presented in \cite{Gomis:2012wy}, we just use the essential points of the prescriptoin. To bring the partition function \eqref{cotangent-flag} into  the LG description, let us define 
\begin{equation}
\Sigma_s^{(I)} = \sigma_s^{(I)} - i\dfrac{B_s^{(I)}}{2 r}~,
\end{equation}
which become the twisted chiral multiplet corresponding to the $I$-th vector multiplet for $\U(I)$. In addition, every ratio of Gamma functions can be replaced by
\begin{equation}
\dfrac{\Gamma(-i r \Sigma)}{\Gamma(1 + i r \overline{\Sigma})} = \int \dfrac{d^2Y}{2\pi}  \text{exp} \Big\{ - e^{-Y} + i r \Sigma Y + e^{-\overline{Y}} + i r \overline{\Sigma} \overline{Y} \Big\} ~,\label{go}
\end{equation}
where $Y$, $\overline{Y}$ represent the twisted chiral fields for the matter sector of the LG model. To study the Coulomb branch of this theory in the infrared, we integrate out the twisted chiral fields  $Y$, $\overline{Y}$. Performing a semiclassical approximation of \eqref{go}
\begin{equation}
Y = - \ln (- i r \Sigma)~,\qquad \overline{Y} = - \ln (i r \overline{\Sigma})~,
\end{equation}
we are left with
\begin{equation}
\dfrac{\Gamma(-i r \Sigma)}{\Gamma(1 + i r \overline{\Sigma})} \,\sim \, \text{exp} \Big\{ \varpi(- i r \Sigma) - \dfrac{1}{2} \ln (-i r \Sigma) -  \varpi(i r \overline{\Sigma}) - \dfrac{1}{2} \ln( i r \overline{\Sigma}) \Big\} ~,\label{exp}
\end{equation}
where $\varpi(x) = x (\ln x - 1)$. This approximation can be also understood as the large radius limit $r\to \infty$ \cite{Benini:2012ui,Bonelli:2014iza}. Using this prescription, we can write 
\bea
Z[T^*\Fl_N] &\sim &\frac{1}{1!\cdots (N-1)!}   \int \prod_{I=1}^{N-1}\prod_{s=1}^{I} \frac{d^2 (r \Sigma_{s}^{(I)})}{2\pi} \Big{\vert}   Q( \Sigma)^{\frac{1}{2}} e^{-\wt \cW_{\text{eff}}(\Sigma)} \Big{\vert}^2 ~,
\eea
where the logarithmic terms in \eqref{exp} give the measure 
\bea
Q(\Sigma) &=  &\prod_{I=2}^{N-1}\prod_{\substack{s,t=1\\ s\neq t}}^{I} (-i r \Sigma_{s}^{(I)}+i r \Sigma_{t}^{(I)}) (-i r \Sigma_{s}^{(I)}+i r \Sigma_{t}^{(I)}+ir\hat m)  \cr
&&\prod_{I=1}^{N-2}\prod_{s=1}^{I}\prod_{u=1}^{{I+1}}  (-i r \Sigma_{s}^{(I)}+i r \Sigma_{u}^{(I+1)})^{-1}(i r \Sigma_{s}^{(I)}-i r \Sigma_{u}^{(I+1)}-ir\hat m)^{-1}\cr
&& \prod_{s=1}^{N-1} (-i r \Sigma_{s}^{(N-1)})^{-1}(i r \Sigma_{s}^{(N-1)}-ir\hat m)^{-1}~,
\eea
and $\wt \cW_{\text{eff}}(\Sigma)$ is the effective twisted superpotential of the mirror LG model in the Coulomb branch
\begin{eqnarray}\label{twisted-superpotential}
\wt \cW_{\text{eff}}(\Sigma)&=&  \sum_{I=1}^{N-1} \sum_{s=1}^{I} (-2\pi\xi^{(I)}+i\theta^{(I)}) (i r \Sigma_s^{(I)} )+ \sum_{I=2}^{N-1} \sum_{s\neq t}^{I} \varpi(-i r \Sigma_{s}^{(I)}+i r \Sigma_{t}^{(I)}+ir\hat m) \cr
&&+ \sum_{I=1}^{N-1} \sum_{s=1}^{I}\sum_{u=1}^{{I+1}}  \left[ \varpi(-i r \Sigma_{s}^{(I)}+i r \Sigma_{u}^{(I+1)}) + \varpi(i r \Sigma_{s}^{(I)}-i r \Sigma_{u}^{(I+1)}-ir\hat m) \right]~.\cr
&&
\end{eqnarray}
where $\Sigma_{s}^{(N)}=a_s$.  
Here we redefine the twisted mass by $m=i\hat m$. Then, the twisted chiral ring is given by the equation of supersymmetric vacua \cite{Nekrasov:2009uh,Nekrasov:2009ui}
\bea\label{vacuum}
\exp\left(\dfrac{\partial \wt \cW_{\text{eff}}}{\partial (i r \Sigma_s^{(I)})}\right)=1~.
\eea
Plugging \eqref{twisted-superpotential} into \eqref{vacuum}, we obtain the following set of equations:
for $I=1$, 
\bea
\prod_{t=1}^{2}\frac{\Sigma_s^{(1)}-\Sigma_t^{(2)}}{\Sigma_s^{(1)}-\Sigma_t^{(2)}-\hat m}= e^{-2\pi\xi^{(1)}+i\theta^{(1)}}~,
\eea
for $1<I<N$, 
\be\label{nested-BAE}
\prod_{t\neq s}^{I}\frac{\Sigma_s^{(I)}-\Sigma_t^{(I)}-\hat m}{\Sigma_s^{(I)}-\Sigma_t^{(I)}+\hat m}\prod_{t=1}^{I-1}\frac{\Sigma_s^{(I)}-\Sigma_t^{(I-1)}+\hat m}{\Sigma_s^{(I)}-\Sigma_t^{(I-1)}}\prod_{t=1}^{I+1}\frac{\Sigma_s^{(I)}-\Sigma_t^{(I+1)}}{\Sigma_s^{(I)}-\Sigma_t^{(I+1)}-\hat m}=\pm e^{-2\pi\xi^{(I)}+i\theta^{(I)}}~.
\ee
These equations are called \emph{nested Bethe ansatz equations} \cite{Krichever:1997,Mukhin:2005} for $\fraksl(N)$ spin chain.  For the cotangent bundle $T^*\textrm{Gr}(r,N)$ of a Grassmannian, the vacuum equation \eqref{vacuum} provides the Bethe ansatz equation of an inhomogeneous $\textrm{XXX}_{\frac{1}{2}}$ spin chain \cite{Nekrasov:2009uh,Nekrasov:2009ui}. Motivated by this physical insight, it was proven in \cite{Maulik:2012wi} that the algebra of quantum multiplication in the equivariant quantum cohomology $QH^*_T(\coprod_{r}T^*\textrm{Gr}(r,N))$ is isomorphic to the maximal commutative subalgebra $B_q$, so-called \emph{Baxter subalgebra}, of Yangain $Y(\fraksl(2))$. Since \eqref{nested-BAE} is the  Bethe ansatz equation for $\fraksl(N)$ spin chain, it is natural to expect that the algebra of quantum multiplication on the equivariant quantum cohomology $QH^*_T(\coprod_{\vec{d}} T^*\Fl(\vec{d}))$ is isomorphic to the Baxter subalgebra of $Y(\fraksl(N))$ \cite{Gorbounov:2013}. (See Appendix \ref{sec:partial} for the definition of a partial flag variety $\Fl(\vec{d})$.) In addition, similar Bethe ansatz equations have been obtained in the system of multiple M2-branes ending on M5-branes \cite{Chen:2013jtk}. It would be interesting to investigate whether there is a duality between the two systems.

\subsubsection{One-loop determinant}\label{sec:1-loop-2*}
The quadratic fluctuations in the $\cN=2^*$ theory receive the contributions from both the vector multiplet and the hypermultiplet in the adjoint representation. Particularly, the one-loop determinant \eqref{hyper-adj-1l} of the hypermultiplet in the adjoint representation can be read off from the index of the Dirac complex tensored with the adjoint bundle. Thus, the one-loop determinant of the $\cN=2^\ast$ theory is expressed as
\be
\scZ_{\textrm{1-loop}}^{\cN=2^\ast}=\prod_{\a\in\Delta}\frac{ \Gamma_2\left(\langle a, \alpha \rangle+m_{\textrm{adj}}+\tfrac{\e_1+\e_2}{2}|\e_1,\e_2\right)\Gamma_2\left(\langle a, \alpha \rangle-m_{\textrm{adj}}+\tfrac{\e_1+\e_2}{2}|\e_1,\e_2\right)}{ \Gamma_2\left(\langle a, \alpha \rangle|\e_1,\e_2\right)\Gamma_2\left(\langle a, \alpha \rangle+\e_1+\e_2|\e_1,\e_2\right)}~.
\ee
Actually, the mass parameter of the hypermultiplet in the instanton partition function is given by $\mu_{\textrm{adj}}=m_{\textrm{adj}} +\frac{\e_1+\e_2}{2}$, and then we can re-write the one-loop determinant with $\mu_{\textrm{adj}}$ in terms of the Upsilon functions:
\bea
\scZ_{\textrm{1-loop}}^{\cN=2^\ast}=\prod_{\a\in\Delta} \frac{\Upsilon\left(\langle a, \alpha \rangle|\e_1,\e_2\right)}{ \Upsilon\left(\langle a, \alpha \rangle+\mu_{\textrm{adj}}|\e_1,\e_2\right)}~.
\eea

With the insertion of a full surface operator, the one-loop determinant can be computed by the same way as in \S\ref{sec:1-loop-pure}. After shifting the equivariant parameters \eqref{equiv shift 2} and taking the $\bZ_N$-invariant part \eqref{average}, we get
\bea\label{1-loop-2*}
\scZ_{\textrm{1-loop}}^{\cN=2^\ast}[1^N]&=&\prod_{i,j=1,i\neq j}^N \tfrac{ \G_2\left(a_i-a_j+\mu_{\textrm{adj}}+\left\lceil\tfrac{j-i}{N}\right\rceil\e_2 |\e_1,\e_2\right) \G_2\left(a_i-a_j -\mu_{\textrm{adj}}+\e_1+\left\lceil\tfrac{1+j-i}{N}\right\rceil\e_2 |\e_1,\e_2\right)}{\G_2\left(a_i-a_j+\left\lceil\tfrac{j-i}{N}\right\rceil\e_2 |\e_1,\e_2\right) \G_2\left(a_i-a_j +\e_1+\left\lceil\tfrac{1+j-i}{N}\right\rceil\e_2 |\e_1,\e_2\right)}\cr
&=&\prod_{i,j=1,i\neq j}^N \frac{\Upsilon\left(a_i-a_j+\left\lceil\tfrac{j-i}{N}\right\rceil\e_2 |\e_1,\e_2\right)}{\Upsilon\left(a_i-a_j+\mu_{\textrm{adj}}+\left\lceil\tfrac{j-i}{N}\right\rceil\e_2 |\e_1,\e_2\right)}~.
\eea
This one loop determinant encodes both 4d and 2d quadratic fluctuations. Indeed, if we subtract the 4d contribution $\scZ_{\textrm{1-loop}}^{\cN=2^\ast}$ from $\scZ_{\textrm{1-loop}}^{\cN=2^\ast}[1^N]$, then we obtain the 1-loop determinant $Z_{\textrm{1-loop}}[T^*\Fl_N]$ of the 2d theory on the surface operator:
\bea\label{1-loop-quotient-2*}
\frac{\scZ_{\textrm{1-loop}}^{\cN=2^\ast}[1^N]}{\scZ_{\textrm{1-loop}}^{\cN=2^\ast}}(a_i,\mu_{\textrm{adj}}=m+\hbar,\e_1=\hbar)
&=&\prod_{\alpha \in \Delta^+} \hbar^{-m-\hbar}\gamma\left(\frac{\langle a, \alpha \rangle}{\hbar}\right)\gamma\left(\frac{-\langle a, \alpha \rangle -m}{\hbar}\right)\cr
&``="&Z_{\textrm{1-loop}}[T^*\Fl_N](a,m,\hbar)~,
\eea
where $``="$ means the equality up to a constant. Here we use the same change of the mass parameter as in \eqref{inst-zero}.

\section{Correlation functions of $\SL(N,\bR)$ WZNW model from gauge theory}\label{sec:3pt}
In a CFT, two-point and three-point functions encode the information about the dynamics of the CFT although the conformal blocks are universal since they are determined by algebras. In the AGT relation, the instanton partition functions coincide with the conformal blocks of the $W_N$/Virasoro algebra, while the one-loop determinants of gauge theory reproduce the product of the three-point functions (the structure constants) of the Toda/Liouviile theory \cite{Dorn:1994xn,Zamolodchikov:1995aa,Fateev:2007ab}. When a full surface operator is inserted, various checks have been carried out for the equivalence between $\SU(N)$ ramified instanton partition functions and  $\wh \fraksl(N)$ conformal blocks \cite{Alday:2010vg,Kozcaz:2010yp,Awata:2010bz,Kanno:2011fw}. The natural candidate for the corresponding part of the one-loop determinants with a full surface operator is the three-point function of $\SL(N,\bR)$ WZNW model. In \S\ref{sec:1-loop-pure} and \S \ref{sec:1-loop-2*}, we have seen the utilities of the orbifold method in the one-loop computations when a full surface operator is present. In this section, we will see that the one-loop determinants of $\SU(2)$ gauge theories computed by the orbifold method indeed reproduce the three-point function of $\SL(2,\bR)$ WZNW model derived in \cite{Teschner:1997ft,Teschner:1999ug,Maldacena:2001km}. Since the three-point function of $\SL(N,\bR)$ WZNW model has not been determined yet, we predict the forms of two-point and three-point function with a semi-degenerate field of $\SL(N,\bR)$ WZNW model by using the one-loop determinants with a full surface operator.

Let us first review the correspondence between a one-loop determinant of a 4d gauge theory without a surface operator and a product of three-point functions of Toda CFT. To this end, we consider the one-loop determinant of  the $\SU(N)$ SCFT with $N_F=2N$. \textit{i.e.} $N$ fundamentals of mass $m_i$ and $N$ anti-fundamentals of $\wt m_i $. The dual correlation function in Toda CFT is the four point function $\langle V_{\b_1}V_{\kappa \omega_{N-1}}V_{\wt\kappa \omega_{N-1}}V_{\wt\b_1}\rangle$ \cite[\S3]{Wyllard:2009hg} with two semi-degenerate fields. Making use of the one-loop determinants \eqref{hyper-general-1l} with redefinitions of the mass parameters 
\bea
\mu_i=-m_i +\frac{\e_1+\e_2}{2} ~, \qquad \wt\mu_i=-\wt m_i -\frac{\e_1+\e_2}{2} ~,
\eea
the one-loop determinant of the $\SU(N)$ SCFT with $N_F=2N$ can be expressed by the product of the Upsilon function
\bea\label{N_F=2N}
\scZ_{\textrm{1-loop}}^{N_F=2N}&=&\frac{ \prod_{\a\in\Delta^+}\Upsilon(\langle a,\alpha \rangle |\e_1,\e_2)\Upsilon(-\langle a,\alpha \rangle |\e_1,\e_2)}{ \prod_{i,j} \Upsilon(\langle a,h_i \rangle  +  \mu_j |\e_1,\e_2) \Upsilon(-\langle a,h_i \rangle -\wt  \mu_j|\e_1,\e_2) }\,,
\eea 
where $h_i$ ($i=1,\cdots,N$) are the weights of the fundamental representation.
On the other hand, the corresponding part of the correlation function in Toda CFT can be determined by the conformal symmetry and $W_N$-symmetry \cite{Fateev:2007ab}. For example,  the reflection amplitude in the two point function of primary fields is expressed as
\begin{equation}\label{MaxRefl-def}
  \langle V_{\beta}(z_1)V_{\beta^{*}}(z_2)\rangle=\frac{R^{-1}(\beta)}{|z_{12}|^{4\Delta(\beta)}}~, \quad   R(\beta)=(\pi\mu\g(b^2))^{\frac{2\langle Q-\b, \a\rangle}{b}}\prod_{\a\in\Delta^+}\tfrac{\G(1+b\langle \b-Q,\a\rangle)\G(b^{-1}\langle \b-Q,\a\rangle)}{\G(1-b\langle \b-Q,\a\rangle)\G(-b^{-1}\langle \b-Q,\a\rangle)}~,
\end{equation}
where $z_{12}=z_1-z_2$, the conformal dimension is given by $\Delta(\beta)=\langle 2Q-\b, \b \rangle/2$, and the conjugated vector parameter $\beta^{*}$ is defined in terms of simple roots $(\alpha_1,\cdots,\alpha_{N-1})\in \Pi$ of $\fraksl(N)$
\begin{equation}\label{Conj-charge}
  (\beta,\a_k)=(\beta^{*},\a_{N-k})~.
\end{equation}
In addition, since the conformal symmetry fixes the form of the three-point functions of primaries \begin{equation*}
\langle V_{\b_1}(z_1)V_{\b_2}(z_2)V_{\b_3}(z_3)\rangle
=\frac{C(\b_1,\b_2,\b_3)} 
      {|z_{12}|^{2\Delta_{12}^3}|z_{13}|^{2\Delta_{13}^2}|z_{23}|^{2\Delta_{23}^1}}\,,
\end{equation*}
where $\Delta_{ij}^k=\Delta(\b_i)+\Delta(\b_j)-\Delta(\b_k)$, it amounts to specifying the structure coefficient $C(\b_1,\b_2,\b_3)$. Although the general structure coefficient in Toda CFT is not known yet, the structure coefficient of the three-point function $\langle V_{\b_1} V_{\varkappa \omega_{N-1}} V_{\b_2}\rangle$ of Toda CFT with a semi-degenerate field $V_{\varkappa \omega_{N-1}}$ \cite{Fateev:2007ab} is given by 
\bea \label{t3pt}
&&    C(\b_1,\varkappa \omega_{N-1},\b_2)\\
&=&    \left[\pi\mu\gamma(b^2)b^{2-2b^2}\right]^{\frac{\langle 2Q-\sum\b_i,\rho \rangle}{b}}\!\!\! \frac{\left(\Upsilon_b(b)\right)^{N-1}\Upsilon_b(\varkappa)\prod_{\a\in \Delta^+}\Upsilon_b\big(\langle Q-\b_1,\a\rangle\big)
    \Upsilon_b\big(\langle Q-\b_2,\a\rangle\big)}{\prod_{ij}\Upsilon_b\big(\frac{\varkappa}{N}+
    \langle\b_1-Q,h_i\rangle+\langle\b_2-Q,h_j \rangle\big)}\,, \nonumber
\eea
where we use the short-hand notation of the Upsilon function for Toda CFT
\bea
\Upsilon_b(x)=\Upsilon(x|b,b^{-1})~.
\eea
Using the shift relation \eqref{eq:updif} of the Upsilon function, one can convince oneself that the reflection amplitude \eqref{MaxRefl-def} can be obtained from the structure coefficient \eqref{t3pt} in the following way:
\bea\label{R-C}
R^{-1}(\b)=C(\b,0,\b^*)~.
\eea
Therefore, the relevant part in the correlation function of Toda CFT can be expressed as
\bea\label{Toda-4pt}
&&C(\b_1,\varkappa \omega_{N-1},\b)R(\b)C(\b^*,\wt\varkappa \omega_{N-1},\wt\b_1)\\
&=&A\frac{\prod_{\a\in \Delta^+}\Upsilon_b\big(\langle Q-\b,\a\rangle\big)
    \Upsilon_b\big(\langle \b-Q,\a\rangle\big)}{\prod_{ij}\Upsilon_b\big(\frac{\varkappa}{N}+
    \langle\b_1-Q,h_i\rangle+\langle\b-Q,h_j \rangle\big )\Upsilon_b\big(\frac{\wt\varkappa}{N}-
    \langle\b-Q,h_i\rangle+\langle\wt\b_1-Q,h_j \rangle\big)}~.\nonumber
\eea
where we confine the unnecessary part to the coefficient $A$. Then, it is easy to see the correspondence between \eqref{N_F=2N} and \eqref{Toda-4pt} upon the identification of the parameters
\bea
a=\b-Q~,\qquad \mu_i=\frac{\varkappa}{N}+
    \langle\b_1-Q,h_i\rangle~,\qquad  \wt \mu_i=-\frac{\wt\varkappa}{N} -\langle\wt\b_1-Q,h_j \rangle~.
\eea

The natural candidate of the 2d CFT dual to $\cN=2$ class $\cS$ theories with a full surface operator is $\SL(N,\bR)$ WZNW model. So far, the two-point and three-point function of $\SL(2,\bR)$ WZNW model are known \cite{Teschner:1997ft,Teschner:1999ug,Maldacena:2001km}. The primary field $ \bV_j(x;z )$ of $\SL(2,\bR)$ WZNW model specified by a highest weight $j$ of the affine Lie algebra $\wh\fraksl(2)$ depends on the isospin coordinate $x$ and the worldsheet coordinate $z$. Then, the two-point function takes the form
\bea
\langle \bV_j(x_1;z_1 )
\bV_{j}(x_2;z_2 ) \rangle=
B(j){ |x_{12}|^{ 4j } \over |z_{12}|^{4\Delta(j)}}
\eea
where $\D(j)=\frac{j(j+1)}{2(k+2)}$ is the conformal dimension of the primary field and the reflection amplitude $B(j)$ is given by
\bea\label{2pt sl2}
B(j) =  -{k+2 \over \pi} { \nu_2^{1+2j} \over \gamma\left(
\tfrac{ 2 j +1 }{ k+2 }\right)}~,\qquad \nu_2 = \pi {\Gamma\left(1 +\tfrac{1}{ k+2} \right)
\over \Gamma\left( 1 -\tfrac{1}{ k+2}\right)}~.
\eea
In addition, the conformal invariance and the affine symmetry determine the three-point function
\bea
\langle \bV_{j_1}(x_1;z_1 )
\bV_{j_2}(x_2;z_2 ) \bV_{j_3}(x_3;z_3 ) \rangle=D(j_1,j_2,j_3)\frac{|x_{12}|^{2j_{12}^3}|x_{13}|^{2j_{13}^2}|x_{23}|^{2j_{23}^1}} 
      {|z_{12}|^{2\Delta_{12}^3}|z_{13}|^{2\Delta_{13}^2}|z_{23}|^{2\Delta_{23}^1}}
\eea
where the structure coefficient $D(j_1,j_2,j_3)$ is given by
\begin{small}
\bea\label{3pt sl2}
&&D( j_3 ,j_2,j_1) \cr
&=&-
\frac{\nu_2^{j_1+j_2+j_3+1} \wt\Upsilon_{k+2}(1) \wt\Upsilon_{k+2}(-2j_1-1)\wt\Upsilon_{k+2}(-2j_2-1)\wt\Upsilon_{k+2}(-2j_3-1)}{2\pi^2\g\left(\tfrac{k+1}{k+2}\right)\wt\Upsilon_{k+2}(-j_1-j_2-j_3-1)\wt\Upsilon_{k+2}(j_3-j_1-j_2)\wt\Upsilon_{k+2}(j_2-j_1-j_3)
\wt\Upsilon_{k+2}(j_1-j_2-j_3)}~.\cr
&&
\eea
\end{small}
Here we use the short-hand notation of the Upsilon function for $\SL(N,\bR)$ WZNW model 
\bea
\wt\Upsilon_{k+N}(x)=\Upsilon (x|1,-k-N)~.
\eea
As in \eqref{R-C}, the reflection amplitude can be obtained from the structure coefficient via 
\be
B(j)=D(j,0,j)~.
\ee

Yet, the two-point and three-point function in $\SL(N,\bR)$ WZNW model are not available even with a semi-degenerate field. It was pointed out in \cite{Kozcaz:2010yp} that, although the primary field $\bV_j(x,z)$ of $\SL(N,\bR)$ WZNW model is dependent of $N(N-1)/2$ isospin variables in general, it suffices to consider only $N-1$ isospin variables $x_i$ ($i=1,\cdots,N-1$) when we deal with the three-point function with a semi-degenerate field. Note that the primary field $\bV_j(\vec{x},z)$ labelled by a highest weight $j$ of $\wh \fraksl(N)$ has its conformal dimension $\D(j)=\frac{\langle j,j+2\rho\rangle}{2(k+N)}$. Besides, the conformal invariance and the affine symmetry constrain the form of the three-point function with a semi-degenerate field \cite[(4.18)]{Kozcaz:2010yp}
\bea
&&\langle \bV_{j_1}(x^{(1)};z_1 )\bV_{j_2=\kappa \omega_{N-1}}(x^{(2)};z_2 ) \bV_{j_3}(x^{(3)};z_3 ) \rangle\cr
&=&\frac{D(j_1,\kappa \omega_{N-1},j_3) } 
      {|z_{12}|^{2\Delta_{12}^3}|z_{13}|^{2\Delta_{13}^2}|z_{23}|^{2\Delta_{23}^1}}\prod_{i=1}^{N-1}|x_i^{(12)}|^{2\langle j_{12}^3,h_i\rangle}|x_i^{(13)}|^{2\langle j_{13}^2,h_i\rangle}|x_i^{(23)}|^{2\langle j_{23}^1,h_i\rangle}~,
\eea
where we define $\langle j_{k\ell}^m,h_i\rangle= \langle j_k+j_\ell-j_m,h_i\rangle$. In the following, let us predict the form of the two-point and three-point function in $\SL(N,\bR)$ WZNW model by making use of the one-loop determinant of the $\SU(N)$ SCFT with $N_F=2N$ in the existence of a full surface operator. Since we have derived the one-loop determinant \eqref{vector HR} of the vector multiplet, we need to determine the one-loop determinant of the hypermultiplet in the (anti-)fundamental representation. As the Coulomb branch parameters are shifted by the holonomy \eqref{equiv shift 2}, we shift the mass parameters due to the orbifold method
\bea\label{mass-shift}
\mu_i \to \mu_i +\frac{N-i}{N}\e_2~,\qquad \wt\mu_i \to \wt\mu_i +\frac{i-1}{N}\e_2~.
\eea
As a result, the one-loop determinant of the hypermultiplet in the fundamental representation is modified as
\bea
&&\prod_{i,j=1}^N\Gamma_2\left(a_i+\mu_j|\e_1,{\e_2}\right)\Gamma_2\left(-a_i-\mu_j+\e_1+\e_2|\e_1,{\e_2}\right)\cr
&\to&\prod_{i,j=1}^N\Gamma_2\left(a_i+\mu_j+\tfrac{N-i-j+1}{N}\e_2|\e_1,\tfrac{\e_2}N\right)\Gamma_2\left(-a_i-\mu_j+\e_1+\tfrac{i+j-N}{N}\e_2|\e_1,\tfrac{\e_2}N\right)~.
\eea
Averaging over the finite group $\bZ_N$ as in \eqref{average}, in the presence of a full surface operator, the one-loop determinant of hypermultiplet in the fundamental representation can be written as 
\bea\label{hyper HR}
\scZ_{\textrm{1-loop}}^{\textrm{hm,fund}}[1^N]&=&\Upsilon\left(a_i+\mu_j+\left\lceil\tfrac{N-i-j+1}{N}\right\rceil   \e_2|\e_1,{\e_2}\right)~.
\eea
After performing the same manipulation for the anti-fundamental representation, the one-loop determinant of the the $\SU(N)$ SCFT with $N_F=2N$ in the presence of a full surface operator can be written as
\bea\label{N_F=2N-2}
&&\scZ_{\textrm{1-loop}}^{N_F=2N}[1^N]\\
&=&\frac{ \prod_{\a\in\Delta^+}\Upsilon(\langle a,\alpha \rangle+\e_2 |\e_1,\e_2)\Upsilon(-\langle a,\alpha \rangle |\e_1,\e_2)}{ \prod_{p,q} \Upsilon\left(\langle a,h_p \rangle +\mu_q+\left\lceil\tfrac{N-p-q+1}{N}\right\rceil \e_2|\e_1,\e_2\right) \Upsilon(-\langle a,h_p \rangle -\wt  \mu_q+\left\lceil\tfrac{p-q}{N}\right\rceil \e_2|\e_1,\e_2) }\,.\nonumber
\eea
When the correspondence between the instanton partition function of the the $\SU(N)$ SCFT with $N_F=2N$ and the $\wh \fraksl(N)$ conformal block part of the four point function $\langle \bV_{j_1}\bV_{\kappa \omega_{N-1}}\bV_{\wt\kappa \omega_{N-1}}\bV_{\wt j_1}\rangle$ was checked in \cite{Kozcaz:2010yp}, the parameters between the 4d gauge theory and  $\SL(N,\bR)$ WZNW model  are identified with\footnote{Here we scale the momenta $j$ and $\kappa$ by two and there are trivial sign differences from  \cite{Kozcaz:2010yp} due to the notation change.}
\be \label{jisa}
\frac{a}{\e_1}=j+\rho\,, \qquad   -\frac{\epsilon_2}{\epsilon_1}= k+N \,,\qquad \frac{{\mu}_i}{\e_1} = -\frac{\varkappa}{N}+\langle  j_1+\rho ,h_i \rangle \,, \qquad  \frac{\wt\mu_{i}}{\e_1} = \frac{\wt\varkappa}{N} - \langle \wt j_1+\rho ,h_{i}\rangle \,. 
\ee 
Using this identification, one can easily deduce the form of three-point function  
\bea\label{3pt slN}
&&D(j_1,\varkappa \omega_{N-1},j_3)\\
&=&A_1\frac{\left(\wt\Upsilon_{k+N}(1)\right)^{N-1}\wt\Upsilon_{k+N}(-\varkappa-1)\prod_{\a\in\Delta^+} \wt\Upsilon_{k+N}(-\langle j_1+\rho,\a \rangle) \wt\Upsilon_{k+N}(-\langle j_3+\rho,\a \rangle)}{\prod_{p,q=1}^N\wt\Upsilon_{k+N}\left(-\frac{\varkappa}{N}+\langle  j_1+\rho , h_q\rangle+\langle j_3+\rho,h_p \rangle  -\left\lceil\tfrac{N-p-q+1}{N}\right\rceil (k+N)\right)}~.\nonumber
\eea
Subsequently, the form of reflection coefficient can be obtained from the three-point function
\be\label{2pt slN}
B(j)=D(j,0,j^*)=\frac{A_2}{\prod_{\a\in\Delta^+}\gamma\left(\tfrac{ \langle 2j+\rho,\a \rangle}{k+N}\right)}~.
\ee
In fact, the relevant part of the four-point correlation function of $\SL(N,\bR)$ WZNW model can be written as
\bea\label{WZNW-4pt}
&&\frac{D( j_1,\varkappa \omega_{N-1},j)D(j^*,\wt\varkappa \omega_{N-1},\wt j_1)}{B(j)}\\
&=&A_3\prod_{\a\in \Delta^+}\wt \Upsilon_{k+N}\big(\langle j+\rho,\a\rangle-(k+N)\big)
    \wt \Upsilon_{k+N}\big(-\langle  j+\rho,\a\rangle\big)\cr
    &&\prod_{p,q}\Bigg[\wt \Upsilon_{k+N}\big(-\tfrac{\varkappa}{N}+ \langle j_1+\rho,h_p\rangle+\langle j+\rho,h_q \rangle    -\left\lceil\tfrac{N-p-q+1}{N}\right\rceil (k+N) \big )\cr
    &&\hspace{1cm}\wt \Upsilon_{k+N}\big(-\tfrac{\wt\varkappa}{N}- \langle j+\rho,h_p\rangle+\langle\wt j_1+\rho,h_q \rangle-\left\lceil\tfrac{p-q}{N}\right\rceil (k+N)\big)\Bigg]^{-1}~,\nonumber
\eea
which is equivalent to \eqref{N_F=2N-2} upon the identification \eqref{jisa} of the parameters. Furthermore, the corresponding part of the one-point correlation function on a torus is equal to
\be
\frac{D(j,\varkappa \omega_{N-1},j^*)}{B(j)}=A_4\prod_{\a\in \Delta^+}\frac{\wt \Upsilon_{k+N}\big(\langle j+\rho,\a\rangle-(k+N)\big)   \wt \Upsilon_{k+N}\big(-\langle  j+\rho,\a\rangle\big)}{\wt \Upsilon_{k+N}\big(-\frac{\varkappa}{N}+\langle j+\rho,\a\rangle-(k+N)\big)
    \wt \Upsilon_{k+N}\big(-\frac{\varkappa}{N}-\langle  j+\rho,\a\rangle\big)}~.
\ee
By the identification \eqref{torus-id} of the parameters, this corresponds to the one-loop determinant \eqref{1-loop-2*} of the $\cN=2^*$ theory. When $N=2$, it is easy to see that \eqref{3pt slN} and \eqref{2pt slN} reduce to \eqref{3pt sl2} and \eqref{2pt sl2}, respectively. This confirms that, when a full surface operator is inserted, a one-loop determinant of an $\SU(2)$ $\cN=2$ gauge theory coincides with a product of the three-point functions of $\SL(2,\bR)$ WZNW model. Nonetheless, the one-loop determinant of the 4d gauge theory cannot determine the coefficients $A_1$ and $A_2$ so that it is important to obtain these coefficients by studying $\SL(N,\bR)$ WZNW model directly.

\section{Discussions}\label{sec:discussion}
The study of the AGT relation with a surface operator that we have implemented raises several questions.
An obvious direction for future work is to study the correlation functions of $\SL(N,\bR)$ WZNW model. Although the gauge theory side has been investigated to some extent, $\SL(N,\bR)$ WZNW model has not been explored at all. In particular, 
the immediate problem left in this paper is to determine the coefficients $A_1$ in \eqref{3pt slN} and $A_2$ in \eqref{2pt slN} of $\SL(N,\bR)$ WZNW model as well as the prefactor $f(t,q)$ in \eqref{prefactor}. It is desirable to obtain a better	comprehension of the effect of the $\cK$ operator.

 In this paper, we study only the pure Yang-Mills and the $\cN=2^*$ theory with a surface operator. The extensive study is needed to provide more complete microscopic descriptions of co-dimension two surface operators in terms of an $\cN=(2,2)$ GLSM coupled to a 4d $\cN=2$ theory as done for co-dimension four surface operators \cite{Gomis:2014eya}. Since the AGT relation tells us that an instanton partition function with a full surface operator obeys a KZ equation, the quantum  connection for  the Higgs branch of the 2d theory on the support of the surface operator can be obtained by a certain limit of the KZ equation. For example,  in the case of the $\SU(N)$ SCFT with $N_F=2N$, the $J$-function of the Higgs branch of the 2d theory should become an eigenfunction of the Painlev\'e VI Hamiltonian \cite{Yamada:2010rr}.

It is intriguing to study K-theoretic $J$-functions \cite{Givental:2003} in terms of $\cN=2$ gauge theories on $S^1\times S^2$. K-theoretic vortex partition functions (a.k.a. holomorphic blocks) \cite{Beem:2012mb,Fujitsuka:2013fga,Benini:2013yva} should compute K-theoretic $J$-functions of the Higgs branches of 3d $\cN=2$ gauge theories. It is well-known that the K-theoretic $J$-function of the complete flag variety becomes an eigenfunction of the $q$-difference Toda operator \cite{Givental:2003}. Recently, it is shown that the K-theoretic $J$-function of the cotangent bundle of the complete flag variety is actually an eigenfunction of  a certain Macdonald difference operator \cite{Braverman:2012}. Hence, it is important to extend these results to the infinite-dimensional version, namely, to find $q$-difference operators of the 5d instanton partition functions with a full surface operator, which should be linked to $q$-KZ equations \cite[\S4.2]{Tan:2013xba}. Besides, it is pointed out in \cite{Kapustin:2013hpk} that the algebra of Wilson loops in 3d $\cN=2$ gauge theory with Chern-Simons term is related to equivariant quantum K-theory of the tautological bundle of a Grassmannian. Further study is required to examine this relationship in order to clarify it.

Another important problem concerns  the relation between co-dimension two and four surface operators. The Liouville correlation functions with appropriate number of degenerate field insertions correspond to $\SL(2,\bR)$ WZNW correlation functions \cite{Ribault:2005wp,Teschner:2010je}, which can be thought of the correspondence between co-dimension two and four surface operators in $\SU(2)$ gauge theories. Nevertheless, the relation in higher rank gauge theories is not understood at all. Since the $W$-algebras are complicated, it would be more amenable to examine the relation by using the microscopic description of surface operators by a coupling of the 4d theories to 2d gauge theories.

%

\section*{Acknowledgement}
The author is indebted to Jaume Gomis for suggesting this project in 2012. Since then, he has benefited through discussion with various people at various occasions. He would like to thank F. Benini, G. Bonelli,  H-Y. Chen, B. Dubrovin, D. Gaiotto, V. Ginzburg, D. Honda, K. Hosomichi, Bumsig K., Hee-Cheol. K., K. Maruyoshi, Sunjay L., V. Pestun, S. Shadrin, R. Suzuki, Y. Tachikawa, A. Tanzini, J. Teschner, P. Vasko, Y. Yamada, for valuable discussions and correspondences. The preliminary versions of the results were presented in ``$\cN=2$  JAZZ workshop 2012'' at McGill University, Algebra and Geometry seminar at University of Amsterdam and the workshop ``Quantum Curves and Quantum Knot Invariants'' at Banff so that the author deeply appreciates J. Seo, G. van der Geer and M. Mulase, respectively, for the kind invitations and their warm hospitality.  He was supported by an STSM Grant from the COST action MP1210 for the stay at SISSA so that he is grateful to both the COST action and SISSA for the support and hospitality. During the stay at SISSA, the result in Appendix \ref{sec:instanton} has been obtained with A. Sciarappa and J. Yagi so that he would like to express his special thanks to them. The work of S.N. is partially supported by the ERC Advanced Grant no.~246974, {\it``Supersymmetry: a window to non-perturbative physics''}.

\appendix
\section{Instanton partition function with surface operator}\label{sec:instanton}
In this appendix, we provide contour integral expressions for the Nekrasov instanton partition function of the chain-saw quiver by making use of the $S^2$ partition functions as done in \cite{Bonelli:2013rja}. The result in this appendix has been obtained with Antonio Sciarappa and Junya Yagi.

A D-brane engineering of the $\cN=2$ $\U(N)$ pure Yang-Mills is provided by a stack of fractional $N$ D3-branes at the 
singular point of the orbifold geometry ${\mathbb C}^2/\bZ_2$. 
The non-perturbative instanton contributions are indeed encoded by  
D(-1)-branes \cite{Witten:1995gx}. In particular, the open string sectors of the D(-1)-D3 system provides the ADHM description of the instanton moduli space where the ADHM constraints  are provided by the D-term and F-term equations.
Hence, the Nekrasov partition function can indeed be computed from the D(-1)-branes point of view as a supersymmetric matrix integral  \cite{Nekrasov:2002qd,Bruzzo:2002xf}.

A more sophisticated description of the construction has been given by resolving the orbifold geometry ${\mathbb C}^2/\bZ_2$ to $T^*S^2$. More specifically, the $\cN=2$ $\U(N)$ pure Yang-Mills is now engineered by $N$ space-time filling D5-branes wrapped on $S^2 \subset T^*S^2$ in Type IIB background $ \bC^2 \times T^*S^2\times \bC$. Now the instanton contributions are encoded by D1-branes wrapped on $S^2 \subset T^*S^2$. From the D1-branes perspective, the D1-D5 system is described by an $\cN=(2,2)$ GLSM on $S^2$ which flows to the NLSM with the instanton moduli space. In fact, the exact partition function of this GLSM computed in \cite{Bonelli:2013rja} captures the $S^2$-finite size corrections to the Nekrasov partition function. Furthermore, it was shown that these corrections encode the equivariant quantum cohomology of the instanton moduli space in terms of Givental $J$-functions. The ordinary instanton partition function can be obtained by taking the zero radius limit of $S^2$.

Although the instanton partition function can be obtained by the D(-1)-D3 system, the D1-D5 system contains richer information. Hence, we shall compute the Nekrasov partition function of the affine Laumon space by using the GLSM description. We consider  Type IIB background on $\bC\times (\bC/\bZ_M)\times T^*S^2\times \bR^2$ with the D1-branes wrapping $S^2$ and spacetime filling D5-branes wrapped on $S^2$. To illustrate the GSLM description of the D1-D5 system, let us briefly recall the chain-saw quiver. The chan-saw quiver $\cM_{\vec{N},\vec{k}}$ is labelled by $\vec{N}=[N_1,N_2,\ldots,N_M]$  and $\vec{k}=[k_1,\cdots,k_M]$ where  the vector spaces $V$ and $W$ are decomposed  according to the representation under the $\bZ_M$ action,
\bea
W = {\bigoplus_{I = 1}^{M}} W_I~, \qquad V = {\bigoplus_{I =1}^{M}} V_I~,
\eea
with
\bea
\dim W_I = N_I~, \qquad  \dim V_I = k_I~. 
\eea
In the language of branes, $W_I$ and $V_I$ are the Chan-Paton spaces of D5- and D1-branes which give rise to $\U(k_I)$ gauge symmetry and $\U(N_I)$ flavor symmetry in the GLSM. Hence, in the chain-saw quiver (Figure \ref{fig:chain-saw}), the linear maps $A_I \in \Hom(V_I,V_I)$ and $B_I  \in \Hom(V_I,V_{I+1})$ are realized from D1-D1 open strings, $P_I \in \Hom(W_I,V_I)$ from D1-D5 open strings and $Q_I \in \Hom(V_I,W_{I+1})$  from D5-D1 open strings. The superpotential of this model is given by $W=\sum_I \Tr_{V_I} \{ \chi_I (A_{I+1} B_I -B_IA_I+P_{I+1}Q_I)\}$ that yields the ADHM equations \eqref{ADHM-eq}. Here the indices $I$ are taken to be modulo $M$. In addition, the equivariant parameters of the torus action $\U(1)^2\times \U(1)^N$ become the twisted masses of the chiral fields. Since the chiral fields $A_I$ and $B_I$ are transformed as the coordinate $z_1$ and $z_2$ \eqref{equiv-quotient} respectively under the spacetime rotation $\U(1)^2$, their twisted masses are given by $-\e_1$ and $-\frac{\e_2}{M}$. It follows from the fact that the superpotential $W$ is trivial under the equivariant action that the chiral fields has the twisted mass $\e=\e_1+\frac{\e_2}{M}$. Because the weight of the equivariant action on $W_I$ is given by the Cartan torus $\U(1)^N$ of $\SU(N)$ with the holonomy shift \eqref{holonomy-shift}, the chiral fields $P_I$ and $Q_{I-1}$ possess the twisted mass $M_{P_{I}}^{(s)} :=-a_{s,I}+\frac{I\e_2}{M}$ and $M_{Q_{I-1}}^{(s)} :=a_{s,I}-\frac{I\e_2}{M}-\e$, respectively. All in all,
 the data about the GLSM is summarized in Table \ref{tab:GLSM}.

\begin{table}[h!]
\begin{center}
\begin{tabular}{c|c|c|c|c|c}
{} & $\chi_I$ & $A_{I}$ & $B_{I}$ & $P_I$ & $Q_{I-1}$ \\ \hline
D-brane sector & D1/D1 & D1/D1 & D1/D1 & D1/D5 & D5/D1 \\ \hline
gauge & $(\mathbf{{k_I}},\mathbf{\bar{k}_{I+1}})$ & $\textrm{\bf Adj}$ & $(\mathbf{\bar{k_I}},\mathbf{k_{I+1}})$ & $\mathbf{k_I}$ & $\mathbf{\bar{k}_{I-1}}$ \\ \hline
flavor  & $\mathbf{1}$ & $\mathbf{1}$ & $\mathbf{1}$ & $\mathbf{\bar{N}}_{I}$ & $\mathbf{N}_{I}$ \\ \hline
twisted mass & $\epsilon=\e_1+\frac{\e_2}{M}$ & $-\epsilon_{1}$ & $-\frac{\epsilon_{2}}{M}$ & $  -a_{s,I} +\frac{I\epsilon_{2}}{M}$ & $ a_{s,I}-\frac{I\epsilon_{2}}{M}-\epsilon$ \\ \hline
$R$-charge & $2$ & $0$ & $0$ & $0$ & $0$ \\ \hline
\end{tabular} 
\caption{Data of GLSM for chain-saw quiver}\label{tab:GLSM}
\end{center}
\end{table}

With these data, it is straightforward to write the Coulomb branch representation of the $S^2$ partition function of the GLSM 
\bea\label{GLSM-chain-saw}
Z[\vec{N},\vec{k};a,\e_1,\e_2] & =& \dfrac{1}{k_1! \ldots k_M!} \sum_{ \substack{ \vec{B}^{(I)} \in \mathbb{Z}^{k_I} \cr I = 1, \ldots , M}} \int \prod_{I=1}^M \prod_{s=1}^{k_I} \dfrac{d (r \sigma_s^{(I)})}{ 2 \pi} e^{- 4 \pi i r \hat{\xi}_I \sigma_s^{(I)} - i \widehat{\theta}_I B_s^{(I)}} \cr
&&\hspace{3cm}\prod_{I=1}^M \prod_{s<t}^{k_I} \left[ (r \sigma_{st}^{(I)})^2 + \dfrac{(B_{st}^{(I)})^2}{4} \right] Z_{\chi_{I}} Z_{A_{I}} Z_{B_{I}} Z_{P_{I}}  Z_{Q_{I}} ~,\cr
Z_{\chi_I} &=& \prod_{I=1}^M \prod_{s=1}^{k_I} \prod_{t=1}^{k_{I+1}}  \tfrac{\Gamma \left( 1 - i r \sigma_s^{(I)} + i r \sigma_t^{(I+1)} - i r \epsilon - \frac{B_s^{(I)}}{2} + \frac{B_t^{(I+1)}}{2} \right)}{\Gamma \left( i r \sigma_s^{(I)} - i r \sigma_t^{(I+1)} + i r \epsilon - \frac{B_s^{(I)}}{2} + \frac{B_t^{(I+1)}}{2} \right)} \cr
Z_{A_I} & =&  \prod_{I=1}^M \prod_{s,t = 1}^{k_I} \tfrac{\Gamma \left( -i r \sigma_{st}^{(I)} + i r \epsilon_1 - \frac{B_{st}^{(I)}}{2} \right)}{\Gamma \left(1 + i r \sigma_{st}^{(I)} - i r \epsilon_1 - \frac{B_{st}^{(I)}}{2} \right)}\cr
Z_{B_I} &=& \prod_{I=1}^M \prod_{s=1}^{k_I} \prod_{t=1}^{k_{I+1}} \tfrac{\Gamma \left( i r \sigma_s^{(I)} - i r \sigma_t^{(I+1)} + i r \frac{\epsilon_2}{M} + \frac{B_s^{(I)}}{2} - \frac{B_t^{(I+1)}}{2} \right)}{\Gamma \left(1 - i r \sigma_s^{(I)} + i r \sigma_t^{(I+1)} - i r \frac{\epsilon_2}{M} + \frac{B_s^{(I)}}{2} - \frac{B_t^{(I+1)}}{2} \right)} \cr
Z_{P_{I}} &=& \prod_{I=1}^M \prod_{s=1}^{k_I}  \prod_{j=1}^{N_I} \tfrac{\Gamma \left( -i r \sigma_s^{(I)} - i r M_{P_{I}}^{(j)} - \frac{B_s^{(I)}}{2} \right)}{\Gamma \left( 1 + i r \sigma_s^{(I)} + i r  M_{P_{I}}^{(j)}- \frac{B_s^{(I)}}{2} \right)}\cr
 Z_{Q_I}  &=&\prod_{I=1}^M \prod_{s=1}^{k_I} \prod_{j=1}^{N_{I+1}} \tfrac{\Gamma \left( i r \sigma_s^{(I)} - i r  M_{Q_{I}}^{(j)}+ \frac{B_s^{(I)}}{2} \right)}{\Gamma \left( 1 - i r \sigma_s^{(I)} + i r M_{Q_{I}}^{(j)} + \frac{B_s^{(I)}}{2} \right)}  ~.
\eea
Writing 
\be
ir\sigma_s^{(I)}=- \frac{B_s^{(I)}}{2}+d_s^{(I)}-ir\phi_s^{(I)}
\ee
we obtain the corresponding Higgs branch formula 
\bea
Z[\vec{N},\vec{k};a,\e_1,\e_2] &=& \frac{1}{k_1! \ldots k_M!} \oint \prod_{I=1}^M \prod_{s=1}^{k_I} \frac{d (i r \phi_s^{(I)}) }{ 2 \pi i}(z_I \bar{z}_I)^{- i r \phi_s^{(I)}} r^{-2ir(N_I - N_{I+1}) \phi_s^{(I)}}  \wt Z_{\text{1-loop}} \wt Z_{\text{v}} \wt Z_{\text{av}}~,\cr
\wt Z_{\text{1-loop}} &=& \left( \tfrac{\Gamma \left( i r \epsilon_1 \right)}{\Gamma \left(1 - i r \epsilon_1 \right)} \right)^{\sum_I k_I} 
\prod_{I=1}^M \prod_{s=1}^{k_I} \prod_{t \neq s}^{k_I} (i r \phi_s^{(I)} - i r \phi_t^{(I)})  \tfrac{\Gamma( i r \phi_s^{(I)} - i r \phi_t^{(I)} + i r \epsilon_1)}{\Gamma(1 - i r \phi_s^{(I)} + i r \phi_t^{(I)} - i r \epsilon_1)}\cr
&&\prod_{I=1}^M \prod_{s=1}^{k_I} \prod_{t=1}^{k_{I+1}} \tfrac{\Gamma \left( - i r \phi_s^{(I)} + i r \phi_t^{(I+1)} + i r \frac{\epsilon_2}{M} \right)}{\Gamma \left(1 + i r \phi_s^{(I)} - i r \phi_t^{(I+1)} - i r \frac{\epsilon_2}{M} \right)}
\tfrac{\Gamma \left( 1 + i r \phi_s^{(I)} - i r \phi_t^{(I+1)} - i r \epsilon \right)}{\Gamma \left( - i r \phi_s^{(I)} + i r \phi_t^{(I+1)} + i r \epsilon \right)} \cr
&&\prod_{I=1}^M \prod_{s=1}^{k_I} \left[ \prod_{j=1}^{N_I} \tfrac{\Gamma \left( i r \phi_s^{(I)} - i r M_{P_I}^{(j)} \right)}{\Gamma \left( 1 - i r \phi_s^{(I)} + i r M_{P_I}^{(j)} \right)}
\prod_{j=1}^{N_{I+1}} \tfrac{\Gamma \left( - i r \phi_s^{(I)} - i r M_{Q_I}^{(j)} \right)}{\Gamma \left( 1 + i r \phi_s^{(I)} + i r  M_{Q_I}^{(j)}  \right)} \right] ~,\cr
\wt Z_{\text{v}} &=& \sum_{ \{ \vec{d} \}} \prod_{I=1}^M \prod_{s=1}^{k_I} \left[ r^{(N_I - N_{I+1})}(-1)^{N_{I+1}}z_I \right]^{d_s^{(I)}} \cr
&&\prod_{I=1}^M \prod_{s < t}^{k_I} \tfrac{d_t^{(I)} - d_s^{(I)} - i r \phi_t^{(I)} + i r \phi_s^{(I)}}{-i r \phi_t^{(I)} + i r \phi_s^{(I)}}  \prod_{s \neq t}^{k_I}( i r \phi_s^{(I)} - i r \phi_t^{(I)} + i r \epsilon_1)_{d_t^{(I)} - d_s^{(I)}}\cr
&&\prod_{I=1}^{M} \prod_{s=1}^{k_I} \prod_{t = 1}^{k_{I+1}} \tfrac{1}{(1 + i r \phi_s^{(I)} - i r \phi_t^{(I+1)} - i r\frac{ \epsilon_2}{M})_{d_t^{(I+1)} - d_s^{(I)}}} \tfrac{1}{(- i r \phi_s^{(I)} + i r \phi_t^{(I+1)} + i r \epsilon)_{d_s^{(I)} - d_t^{(I+1)} }}\cr
&&\prod_{I=1}^{M} \prod_{s=1}^{k_I}  \tfrac{\prod_{j=1}^{N_{I+1}} (- i r \phi_s^{(I)} - i r M_{Q_{I}}^{(j)})_{d_s^{(I)}}}{\prod_{j=1}^{N_I}(1 - i r \phi_s^{(I)} + i r  M_{P_{I}}^{(j)})_{d_s^{(I)}}} ~,\cr
\wt Z_{\text{av}} &=& \sum_{ \{ \vec{\tilde{d}} \}} \prod_{I=1}^M \prod_{s=1}^{k_I} \left[ r^{(N_I - N_{I+1})}(-1)^{N_{I+1}}\overline z_I \right]^{\tilde{d}_s^{(I)}} \cr
&&\prod_{I=1}^M \prod_{s < t}^{k_I} \tfrac{\tilde{d}_t^{(I)} - \tilde{d}_s^{(I)} - i r \phi_t^{(I)} + i r \phi_s^{(I)}}{-i r \phi_t^{(I)} + i r \phi_s^{(I)}}  \prod_{s \neq t}^{k_I} ( i r \phi_s^{(I)} - i r \phi_t^{(I)} + i r \epsilon_1)_{\tilde{d}_t^{(I)} - \tilde{d}_s^{(I)}}\cr
&&\prod_{I=1}^{M} \prod_{s=1}^{k_I} \prod_{t = 1}^{k_{I+1}} \tfrac{1}{(1 + i r \phi_s^{(I)} - i r \phi_t^{(I+1)} - i r\frac{ \epsilon_2}{M})_{\tilde{d}_t^{(I+1)} - \tilde{d}_s^{(I)}}} \tfrac{1}{(- i r \phi_s^{(I)} + i r \phi_t^{(I+1)} + i r \epsilon)_{\tilde{d}_s^{(I)} - \tilde{d}_t^{(I+1)} }}\cr
&&\prod_{I=1}^{M} \prod_{s=1}^{k_I}  \tfrac{\prod_{j=1}^{N_{I+1}} (- i r \phi_s^{(I)} - i r M_{Q_{I}}^{(j)})_{\tilde{d}_s^{(I)}}}{\prod_{j=1}^{N_I}(1 - i r \phi_s^{(I)} + i r  M_{P_{I}}^{(j)})_{\tilde{d}_s^{(I)}}} ~.
\eea
Note that $\wt Z_v$ can be interpreted as the $J$-function of the affine Laumon space.
In the zero radius limit $r \rightarrow 0$, the partition function receives the contribution only from $\wt Z_{\text{1-loop}}$, leaving the generating function of the equivariant cohomology of the chain-saw quiver $\cM_{\vec{N},\vec{k}}$ 
\bea	\label{pure-inst-so}
\scZ_{\vec{N},\vec{k}}^{\rm pure} &=& \prod_{I=1}^M \dfrac{1}{ k_I! (2 \pi i \epsilon_1)^{k_I}} \oint  \prod_{I=1}^M \prod_{s=1}^{k_I} \dfrac{d \phi_s^{(I)}}{ \prod_{j=1}^{N_I}(\phi_s^{(I)} - M_{P_I}^{(j)}) \prod_{j=1}^{N_{I+1}}(\phi_s^{(I)} +M_{Q_I}^{(j)})} \cr
&&\hspace{2.5cm} \prod_{I=1}^M \prod_{s=1}^{k_I} \prod_{t \neq s}^{k_I} \dfrac{ \phi_{st}^{(I)} }{\phi_{st}^{(I)} + \epsilon_1} 
\prod_{I=1}^M \prod_{s=1}^{k_I} \prod_{t=1}^{k_{I+1}} \dfrac{\phi_{s}^{(I)} - \phi_t^{(I+1)} + \epsilon}{\phi_{s}^{(I)} - \phi_t^{(I+1)} + \frac{\epsilon_2}{M}}~.
\eea
The poles of this contour integral are classified by the $N$-tuple of Young diagrams $\vec{Y}=(Y^{s,I})$ $(I=1,\cdots,M, \ s=1,\cdots,N_I)$ where the boxes in the $j$-th column of $Y^{s,I}$ contribute to the instanton number $k_{I+j-1}$.
We verify that the residues match with the result \cite[Mathematica file]{Kanno:2011fw} in various values of $(\vec{N},\vec{k})$. 

Furthermore, since the $\cN=4$ ADHM data is given \cite[\S2.1]{Bruzzo:2002xf} \cite[X.3.1]{Dorey:2002ik}, one can derive the instanton partition function with the surface operator for the $\cN=2^\ast$ theory in a similar manner. For brevity, we just present the final result:
\bea\label{N=2*-general}
\scZ_{\vec{N},\vec{k}}^{\cN=2^\ast} &=& \prod_{I=1}^M \dfrac{(\e_1-\mu_{\textrm{adj}})^{k_I}}{ k_I! (2 \pi i \epsilon_1\mu_{\textrm{adj}})^{k_I}} \cr
&&\oint  \prod_{I=1}^M \prod_{s=1}^{k_I}d \phi_s^{(I)} \dfrac{ \prod_{j=1}^{N_I}(\phi_s^{(I)} - M_{P_I}^{(j)}+\mu_{\textrm{adj}}) \prod_{j=1}^{N_{I+1}}(\phi_s^{(I)} +M_{Q_I}^{(j)}-\mu_{\textrm{adj}})}{ \prod_{j=1}^{N_I}(\phi_s^{(I)} - M_{P_I}^{(j)}) \prod_{j=1}^{N_{I+1}}(\phi_s^{(I)} + M_{Q_I}^{(j)})} \cr
&& \prod_{I=1}^M \prod_{s=1}^{k_I} \prod_{t \neq s}^{k_I} \dfrac{ \phi_{st}^{(I)} (\phi_{st}^{(I)} + \epsilon_1-\mu_{\textrm{adj}})}{(\phi_{st}^{(I)}+ \mu_{\textrm{adj}})(\phi_{st}^{(I)} + \epsilon_1)} \cr
&&\prod_{I=1}^M \prod_{s=1}^{k_I} \prod_{t=1}^{k_{I+1}} \dfrac{(\phi_{s}^{(I)} - \phi_t^{(I+1)} + \epsilon)(\phi_{s}^{(I)} - \phi_t^{(I+1)} + \frac{\epsilon_2}{M}-\mu_{\textrm{adj}})}{(\phi_{s}^{(I)} - \phi_t^{(I+1)} + \frac{\epsilon_2}{M})(\phi_{s}^{(I)} - \phi_t^{(I+1)} + \epsilon-\mu_{\textrm{adj}})}~.
\eea

Let us conclude this appendix by mentioning a relation between quantum cohomology of the affine Laumon space and  quantum integrable system. It was found in \cite{Bonelli:2014iza} that there is the relation between the $\frakgl(N)$ intermediate long wave integrable system and the quantum cohomology of the ADHM instanton moduli space. More precisely, the authors of \cite{Bonelli:2014iza} shows that the effective twisted superpotential in the Landau-Ginzburg mirror of the GLSM with the standard ADHM instanton moduli space coincides with the Yang-Yang potential of the $\frakgl(N)$ intermediate long wave integrable system \cite{Litvinov:2013zda}.

Thus, to see quantum integrable structure behind the quantum cohomology of the affine Laumon space, we can perform the same analysis done in \S \ref{sec:bethe}. By defining
\bea
\Sigma_s^{(I)}\equiv\sigma_s^{(I)}-i\frac{B_s^{(I)}}{2r}~,
\eea
we can obtain the effective twisted superpotential of the Landau-Ginzburg mirror of the chain-saw quiver by taking the large radius limit of \eqref{GLSM-chain-saw}:
\bea
Z[\vec{N},\vec{k};a,\e_1,\e_2] & \sim & \dfrac{1}{k_1! \ldots k_M!} \int   \prod_{I=1}^M \prod_{s=1}^{k_I}\frac{d^2 \Sigma_{s}^{(I)}}{2\pi} \Big{\vert}   Q( \Sigma)^{\frac{1}{2}} e^{-\wt \cW_{\text{eff}}} \Big{\vert}^2 ~,
\eea
where the measure is written as
\bea
 Q&=&\prod_{I=1}^M \prod_{\substack{s,t = 1\\ s\neq t}}^{k_I}  \prod_{u=1}^{k_{I+1}}\prod_{j=1}^{N_I} \prod_{\ell=1}^{N_{I+1}}  \frac{(\Sigma_{st}^{(I)})(\Sigma_s^{(I)}-\Sigma_u^{(I+1)}+\e)}{(\Sigma_{st}^{(I)}-\e_1)(\Sigma_s^{(I)}-\Sigma_u^{(I+1)}+\tfrac{\e_2}{M}) (\Sigma_s^{(I)}+M_{P_{I}}^{(j)})(\Sigma_s^{(I)}-M_{Q_{I}}^{(\ell)})}~,\cr
 &&
\eea
and the effective twisted superpotential is given by
\bea\label{twisted-sp-chainsaw}
\wt \cW_{\text{eff}}&=& \sum_{I=1}^M \sum_{s=1}^{k_I}\Bigg[ -(2\pi\xi^{(I)}-i\theta^{(I)})(ir\Sigma_s^{(I)})\\
&&+\sum_{u=1}^{k_{I+1}} -\varpi\left(ir(\Sigma_s^{(I)}-\Sigma_u^{(I+1)}+\e)\right) +\varpi\left(ir(\Sigma_s^{(I)}-\Sigma_u^{(I+1)}+\tfrac{\e_2}{M})\right)\cr
&&+\sum_{t = 1}^{k_I}\varpi\left(-ir(\Sigma_{st}^{(I)}-\e_1)\right)+\sum_{j=1}^{N_I}\varpi\left(-ir(\Sigma_s^{(I)}+M_{P_{I}}^{(j)})\right)+\sum_{(\ell=1}^{N_{I+1}}\varpi\left(ir(\Sigma_s^{(I)}-M_{Q_{I}}^{\ell)})\right)\Bigg]~.\nonumber
\eea
It would be interesting to find the quantum integrable system whose Yang-Yang potential coincides with \eqref{twisted-sp-chainsaw}. For instance, in the case of $N=2$ and $[1,1]$ partition, the vacuum equation 
\be
\exp\Bigg( \frac{\partial \wt \cW_{\text{eff}}}{\partial (i r \Sigma_s^{(I)})}\Bigg)=1~,
\ee
leads to the Bethe equation 
\begin{small}
\bea
\prod_{t\neq s}^{k_1} \frac{(\Sigma_s^{(1)}-\Sigma_t^{(1)}-\e_1)}{(\Sigma_s^{(1)}-\Sigma_t^{(1)}+\e_1)}\prod_{t=1}^{k_2}\frac{(\Sigma_s^{(1)}-\Sigma_t^{(2)}-\frac{\e_2}{2})(\Sigma_s^{(1)}-\Sigma_t^{(2)}+\e)}{(\Sigma_s^{(1)}-\Sigma_t^{(2)}+\frac{\e_2}{2})(\Sigma_s^{(1)}-\Sigma_t^{(2)}-\e)}&=&\pm e^{-2\pi\xi^{(1)}+i\theta^{(1)} }\frac{(\Sigma_s^{(1)}-M_{Q_1})}{(\Sigma_s^{(1)}+M_{P_1})}\cr
\prod_{t\neq s}^{k_2} \frac{(\Sigma_s^{(2)}-\Sigma_t^{(2)}-\e_1)}{(\Sigma_s^{(2)}-\Sigma_t^{(2)}+\e_1)}\prod_{t=1}^{k_1}\frac{(\Sigma_s^{(2)}-\Sigma_t^{(1)}-\frac{\e_2}{2})(\Sigma_s^{(2)}-\Sigma_t^{(1)}+\e)}{(\Sigma_s^{(2)}-\Sigma_t^{(1)}+\frac{\e_2}{2})(\Sigma_s^{(2)}-\Sigma_t^{(1)}-\e)}&=&\pm e^{-2\pi\xi^{(2)}+i\theta^{(2)}  }\frac{(\Sigma_s^{(2)}-M_{Q_2})}{(\Sigma_s^{(2)}+M_{P_2})}~.\cr
&&
\eea
\end{small}
This can be interpreted as the spin version of the Bethe ansatz equation for the intermediate long wave integrable system \cite{Bonelli:2014iza,Litvinov:2013zda}.

\section{One-loop determinants}\label{sec:1-loop}

In this appendix, we shall elaborate the computation of one-loop determinants on the orbifold space $\bC\times (\bC/\bZ_N)$.
We start with a brief review of the one-loop computations using the Atiyah-Singer equivariant index theorem.
For more detail, we refer the reader to \cite{Pestun:2007rz,Gomis:2011pf,Hama:2012bg}.

The exact partition functions of $\cN=2$ supersymmetric Yang-Mills theories on $S^4_b$ can be evaluated by applying supersymmetric localization. The value of an infinite-dimensional functional integral is invariant under the deformation $S \to S+t \hat Q \hat V$ of the action $S$ by a $\hat Q$-exact term where $\hat Q=Q+Q_{\textrm{BRST}}$ is the combination of a supercharge and a BRST operator and $\hat V=V+V_{\textrm{ghost}}$ is the combination of $V=(\Psi,\overline{Q\Psi})$ and the gauge fixing term $V_{\textrm{ghost}}$. In the limit of  $t \to \infty$, the term ${t \hat Q \hat V}$ dominates in the infinite-dimensional functional integral, which renders the one-loop approximation at the BPS configurations $\hat Q \hat V=0$:
\bea
\scZ=\int_{\hat Q \hat V=0} \scZ_{\textrm{1-loop}} ~, \qquad  \scZ_{\textrm{1-loop}}= \left[\frac{\det K_{\textrm{fermion}}}{\det K_{\textrm{boson}}}\right]^{\frac12}~,
\eea
where $K_{\textrm{boson}}$ and  $K_{\textrm{fermion}}$ are the kinetic operators of 
\be
\hat Q \hat V=(X_{\textrm{boson}},K_{\textrm{boson}}X_{\textrm{boson}})+(X_{\textrm{fermion}},K_{\textrm{fermion}}X_{\textrm{fermion}})~.
\ee
In this one-loop determinant, there occurs the cancellation between the bosonic and the fermionic fluctuations when they are paired by the supercharge $Q$. Hence it receives the contribution only from the kernel and cokernel spaces of the transversal elliptic operator $D$ that is the quadratic operator in $\hat V$ so that 
\be
\scZ_{\textrm{1-loop}}= \left[\frac{\det_{\textrm{Coker}D} \cR}{\det_{\textrm{Ker}D} \cR}\right]^{\frac12}~,
\ee
where $\hat Q^2=\cR$ is the generator of the product $\SO(4)\times \SU(N)\times G_F$ of the spacetime, guage and flavor symmetry. Therefore, the one-loop determinants can be obtained by the product of weights for the group action
$\cR$ on the kernel and cokernel spaces of $D$. This is encoded in the $\cR$-equivariant index
\be
{\rm ind}\,D={\rm tr}_{{\rm Ker}D} e^\cR-{\rm tr}_{{\rm Coker}D} e^\cR\ ,
\label{indexcharacter}
\ee
which can then be calculated from the equivariant Atiyah-Singer index theorem \cite{Atiyah:1974}. Since the index $\text{ind}\,D$  is expressed as the sum over weights, we can convert the index into the determinant via
\begin{eqnarray}
\label{rule-appA}
\sum_j c_j e^{w_j(\e_1,\e_2,a,m_f)}
\rightarrow \prod_j w_j(\e_1,\e_2,a,m_f)^{c_j}\,,
\end{eqnarray}
where $(\e_1,\e_2,a,m_f)$ denote the equivariant parameters for $\SO(4)\times \SU(N)\times G_{\text F}$.

For $\cN=2$ supersymmetric gauge theories $S^4_b$, the critical points $\hat Q \hat V=0$ consist of self-dual connections $F^+=0$ at the north pole and anti-self-dual connections $F^-=0$ at the south pole so that we consider the equivariant index around these configurations \cite{Pestun:2007rz}. Let us first compute the index for the vector multiplet. Near the north pole,  the operator $D^{\textrm{vm}}$ for the vector mutiplet is actually the complex  of vector bundles associated with linearization of the self-dual
equation $F^{+} = 0$ on $\bR^{4}$
\begin{equation}
\label{eq:self-dual}
 D_\text{SD}:    \Omega^{0} \stackrel{d}{\to} \Omega^{1} \stackrel{d_{+}}{\to} \Omega^{2+}\,.
\end{equation}
where $d_{+}$ is the composition of
the de Rham differential and self-dual projection operator. Then, tensoring the adjoint
representation  of the gauge group with this complex, the  $\U(1)^2 \times
\U(1)^N$-equivariant index for the vector multiplet can be computed by the Atiyah-Singer index theorem \cite{Atiyah:1974} in a simple way
\begin{equation}
  \ind (D^\vm)(\e_1, \e_2, a) =
 \frac{ (1 + e^{ i \e_1 + i \e_2})}{ (1-e^{i \e_1}
    )(1-e^{i \e_2})} \sum_{w\in\text{adj}} e^{ i \langle a, w \rangle }\,.
    \label{newindex}
\end{equation}
where $w$ is a weight of the adjoint representation of $\SU(N)$. At the south pole, we expand   \eqref{newindex}  in terms of negative powers of $e^{ i \e_1}$ and $e^{ i \e_2}$, which results in the sign change $(\e_1,\e_2) \to (-\e_1,-\e_2)$. This can be absorbed into the reflection of weights $w \to -w$. Hence, it gives rise to the identical contribution to the one-loop determinant. Then, using \eqref{rule-appA}, one can write the one-loop determinant of the vector multiplet  
\be\label{vector-1l}
\scZ_{\textrm{1-loop}}^{\textrm{vm}}=\prod_{\a\in\Delta} \left[\Gamma_2\left(\langle a, \alpha \rangle|\e_1,\e_2\right)\Gamma_2\left(\langle a, \alpha \rangle+\e_1+\e_2|\e_1,\e_2\right)\right]^{-1}~,
\ee
where the Barnes double Gamma function $\Gamma_2(x|\e_1,\e_2)$ can be considered as the regularized infinite product
\be
\Gamma_2(x|\e_1,\e_2)\propto \prod_{n,m=0}^\infty(x+m\e_1+n\e_2)^{-1}~.
\ee
The precise definition of the Barnes double Gamma function $\Gamma_2(x|\e_1,\e_2)$ is given in the end of this section.

Next, we shall evaluate the hypermultiplet contribution to the one-loop determinant.
The transversal elliptic operator $D^{\textrm{hm}}$ for a hypermultiplet is the Dirac operator $D_{\text{Dirac}}$
that maps the spinor bundle $S^{+}$ of positive-chirality 
 to the spinor bundle $S^{-}$ of negative-chirality
\begin{equation}
  \label{eq:Scomplex}
D_{\text{Dirac}}: S^{+} \to S^{-}\,.
\end{equation}
An equivariant index for a hypermultiplet depends on the representation of the gauge group. For  a hypermultiplet in the adjoint
representation, the Dirac complex is tensored with the adjoint bundle on which the $G_{\text F}=\SU(2)$ flavor symmetry acts on.
Therefore the 
  $\U(1)^2 \times
\SU(N)\times G_{\text F}$ equivariant index  is given by
\begin{equation}
  \ind D^\hm_{\textrm{adj}} (\e_1,\e_2,a, m_{\textrm{adj}}) 
 =  -  \frac { 
e^{\frac 1 2 (i \e_1 + i \e_2) }
}
 {(1 -e^{i\e_1})(1-e^{i\e_2})}(e^{i m_{\textrm{adj}}} + e^{-i m_{\textrm{adj}}})
\sum_{w\in \text{adj}} e^{i \langle a, w \rangle}\,.
\label{eq:index-adj}
\end{equation}
where $m_{\textrm{adj}}$ is  the equivariant parameter of the $\SU(2)$ flavor symmetry. Since the contribution from the south pole is the same as that from the north pole, the one-loop determinant of  a hypermultiplet in the adjoint
representation is given by
\be\label{hyper-adj-1l}
\scZ_{\textrm{1-loop}}^{\textrm{hm, adj}}=\prod_{\a\in\Delta} \Gamma_2\left(\langle a, \alpha \rangle+m_{\textrm{adj}}+\tfrac{\e_1+\e_2}{2}|\e_1,\e_2\right)\Gamma_2\left(\langle a, \alpha \rangle-m_{\textrm{adj}}+\tfrac{\e_1+\e_2}{2}|\e_1,\e_2\right)~.
\ee

The equivariant index for a hypermultiplet in an arbitrary representation $R$ of the gauge group is rather subtle since there occurs an enhancement of a flavor group in some representations. We refer the reader to  \cite{Gomis:2011pf} in which the detail analysis is provided.
In conclusion, for a hypemultiplet in an arbitrary  representation $R$, the $\U(1)^2\times\SU(N)\times G_{\text F}$-equivariant index can be expressed as
\begin{equation}
  \ind D^\hm_{R}(\e_1, \e_2,a,m_f) 
 =  -  \frac { 
e^{\frac 1 2 (i \e_1 + i \e_2) }
}
 {(1 -e^{i\e_1})(1-e^{i\e_2})}
\sum_{f=1}^{N_\text{F}}
\sum_{w\in R} \left(
e^{i \langle a, w \rangle-i m_f}
+e^{-i \langle a, w \rangle+i m_f}
\right)
\,.
\label{eq:index-univ}
\end{equation}
where $N_\text{F}$ mass parameters $m_f$ with $f=1,\ldots N_\text{F}$  parametrizes 
the Cartan subalgebra of $G_{\text F}$.
Therefore, the one-loop determinant of a hypermultiplet in a representation $R$ can be expressed as
\be\label{hyper-general-1l}
\scZ_{\textrm{1-loop}}^{\textrm{hm}~R}=\prod_{f=1}^{N_\text{F}} \prod_{w\in R} \Gamma_2\left(\langle a, w \rangle-m_{f}+\tfrac{\e_1+\e_2}{2}|\e_1,\e_2\right)\Gamma_2\left(-\langle a, w \rangle+m_{f}+\tfrac{\e_1+\e_2}{2}|\e_1,\e_2\right)~.
\ee

Since the instanton partition functions with a full surface operator can be obtained by applying the localization method to the instanton moduli space on the orbifold space $\bC\times (\bC/\bZ_N)$, it is reasonable to expect that the one-loop determinant can be also computed by the orbifold procedure. Due to the orbifold space $\bC\times (\bC/\bZ_N)$, we need to take the fractional equivariant parameter $\e_2\to\frac{\e_2}{N}$ \eqref{equiv-quotient}, and the coulomb \eqref{holonomy-shift} and mass parameters \eqref{mass-shift} with holonomy shift. Hence, the part of a one-loop determinant that takes the form $\Gamma_2(x|\e_1,\e_2)$ on $\bC^2$ is generally altered in the following way:
\bea\label{one-loop-shift}
\Gamma_2(x(a,m_f,\e_1,\e_2)|\e_1,\e_2)\rightarrow \Gamma_2\left(\tilde x(a,m_f,\e_1)+\tfrac{I\e_2}{N}\Big|\e_1,\tfrac{\e_2}{N}\right)
\eea
Then, its $\bZ_N$-invariant part becomes the one-loop determinant on  $\bC\times (\bC/\bZ_N)$. To take the $\bZ_N$-invariant part, it is easy to use the index. Writing $t=e^{i\e_2/N}$, the index that corresponds to the right hand side of \eqref{one-loop-shift} is
\bea
g(t)=e^{i\tilde x} \frac{ t^{I}}{1- t} ~.
\eea
The  $\bZ_N$-invariant part can be taken by averaging over the $\bZ_N$ group 
\bea\label{average}
\frac1N\sum_{k=0}^{N-1}g(\omega^k t)=e^{i\tilde x}~ \frac{t^{\lceil \frac{I}{N} \rceil N}}{1-t^N}~,
\eea
where $\omega=\exp(2\pi i/N)$ is the $N$-th root of unity and   $\lceil x\rceil$ denotes the smallest integer $\ge x$. Subsequently, the one-loop determinant on $\bC\times (\bC/\bZ_N)$ can be written as
\bea
\scZ_{\textrm{1-loop}}[\bC\times (\bC/\bZ_N)]= \Gamma_2\Big(\tilde x(a,m_f,\e_1)+\left\lceil\tfrac{I\e_2}{N} \right\rceil \Big|\e_1,\e_2 \Big) ~.
\eea
For concrete illustration, let us show simple examples in the case of  $\bC\times (\bC/\bZ_2)$. The fractional equivariant parameter $\frac{\e_2}{2}$ and the holonomy shift generally ends up with the Barnes double gamma function $\Gamma_2(x|\e_1,\frac{\e_2}{2})$ whose pole structure is depicted in Figure \ref{fig:lattice}. Roughly speaking, we need to take the even modes from them.  For instance, the even modes can be read off by averaging over the $\bZ_2$ group
\bea\label{even mode}
\G_2\left(x+\e_2|\e_1,\tfrac{\e_2}2\right) \to\G_2\left(x+\e_2|\e_1,\e_2\right) &&\quad \frac12\left[\frac{t^2}{1-t}+ \frac{(-t)^2}{1-(-t)}\right]=\frac{t^2}{1-t^2}\cr
\G_2\left(x+\tfrac{\e_2}2|\e_1,\tfrac{\e_2}2\right) \to\G_2\left(x+\e_2|\e_1,\e_2\right) &&\quad \frac12\left[\frac{t}{1-t}+ \frac{(-t)}{1-(-t)}\right]=\frac{t^2}{1-t^2} \cr
\G_2\left(x|\e_1,\tfrac{\e_2}2\right) \to\G_2\left(x|\e_1,\e_2\right)  &&\quad \frac12\left[\frac{1}{1-t}+ \frac{1}{1-(-t)}\right]=\frac{1}{1-t^2}\cr
\G_2\left(x-\tfrac{\e_2}2|\e_1,\tfrac{\e_2}2\right) \to\G_2\left(x|\e_1,\e_2\right)  &&\quad \frac12\left[\frac{t^{-1}}{1-t}+ \frac{(-t^{-1})}{1-(-t)}\right]=\frac{1}{1-t^2}~.
\eea

\begin{figure}[h]
 \centering
 \includegraphics[width=5.5cm]{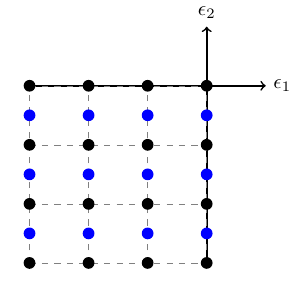}
    \caption{The distribution of poles of $\Gamma_2(x|\e_1,\frac{\e_2}{2})$. Only poles with black color are $\bZ_2$-invariant. }\label{fig:lattice}
\end{figure}

Let us conclude this section by providing the definitions of the special functions that appear in this paper.
The Barnes double Gamma function $\Gamma_2(x|\e_1,\e_2)$ is defined by 
\bea\label{Barnes}
\Gamma_2(x|\e_1,\e_2):=\exp\left[
\frac{d}{ds}\bigg|_{s=0}\zeta_2(s;x|\e_1,\e_2)
\right]\,,
\eea
where the double zeta function is provided as
\bea
\zeta_2(s;x|\e_1,\e_2)
=\sum_{m,n}(m\e_1+n\e_2+x)^{-s}
=\frac{1}{\Gamma(s)}\int_0^\infty
\frac{dt}{t}
\frac{t^s e^{-tx}}{(1-e^{-\e_1t})(1-e^{-\e_2t})}\,.
\eea
In this paper, we also use the Upsilon function which is the product of the Barnes double Gamma functions
\begin{equation} \label{Upsilon}
\Upsilon(x|\e_1,\e_2) := \frac{1}{\Gamma_2(x|\e_1,\e_2)\Gamma_2(\e_1+\e_2-x|\e_1,\e_2)}~,
\end{equation}
and therefore it obeys
\bea
\Upsilon(x|\e_1,\e_2)=\Upsilon(\e_1+\e_2-x|\e_1,\e_2)~.
\eea
Besides, it admits the following line integral representation
\begin{equation}
\log\Upsilon(x|\e_1,\e_2) = \int_0^\infty \frac{dt}{t}\left[\frac{(\e_1+\e_2-2x)^2}{4}e^{-2t}-\frac{\sinh^2(\e_1+\e_2-2x)\frac{t}{2}}{\sinh(\e_1 t)\sinh(\e_2 t)}\right].
\end{equation}
The characteristic property of the Upsilon function is the shift relation
\bea
\Upsilon(x+\e_1|\e_1,\e_2) &=&\e_2^{2x/\e_2-1}\g(x/\e_2)\Upsilon(x|\e_1,\e_2) \cr
\Upsilon(x+\e_2|\e_1,\e_2) &= &\e_1^{2x/\e_1-1}\g(x/\e_1)\Upsilon(x|\e_1,\e_2) ~, \label{eq:updif}
\eea
which plays an important role in this paper.

\section{$J$-function of cotangent bundle of partial flag variety}\label{sec:partial}
\begin{figure}[h]
 \centering
 \includegraphics[width=10cm]{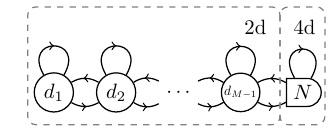}
    \caption{Quiver diagram for $\cN=(2,2)^*$ GLSM whose Higgs branch is the cotangent bundle $T^*\Fl(\vec{d})$ of a partial flag variety.}\label{fig:partial}
\end{figure}
The partial flag variety $\Fl(\vec{d})=\Fl(d_1,\cdots,d_{M-1},d_M=N)$ is an increasing sequence of linear subspaces of $\bC^N$
\bea
0\subset \bC^{d_1}\subset \cdots \subset \bC^{d_{M-1}}\subset  \bC^{d_{M}}=\bC^N~.
\eea
Thus, the GLSM given in Figure \ref{fig:partial} flows to NLSM with $T^\ast\Fl(\vec{d})$. As in \S\ref{sec:cotangent}, one can extract the $J$-function of $T^\ast\Fl(\vec{d})$ from the $S^2$ partition function of the GLSM:
\bea
J[T^*\Fl(\vec{d})]  &=& \sum_{\vec{k}^{(I)}}  \prod_{I=1}^{M-1}z_I^{\vert k^{(I)}\vert } \prod_{I=1}^{M-1}\prod_{s\neq t}^{d_I} \tfrac{(1+\hbar^{-1} H_{st}^{(I)}+\hbar^{-1}m)_{k_s^{(I)}-k_t^{(I)}}}{(\hbar^{-1} H_{st}^{(I)})_{k_s^{(I)}-k_t^{(I)}}} \cr
&& \prod_{I=1}^{M-2} \prod_{s=1}^{d_I}\prod_{t=1}^{d_{I+1}} \tfrac{(\hbar^{-1} H_{s}^{(I)} -\hbar^{-1} H_{t}^{(I+1)}-\hbar^{-1}m)_{k^{(I)}_s - k^{(I+1)}_t}}{(1+\hbar^{-1} H_{s}^{(I)} -\hbar^{-1} H_{t}^{(I+1)})_{k^{(I)}_s - k^{(I+1)}_t}}  \cr 
&&\prod_{s=1}^{d_{M-1}} \prod_{t=1}^{N} \tfrac{(\hbar^{-1} H_{s}^{(M-1)}-\hbar^{-1} H_{t}^{(M)}-\hbar^{-1}m)_{k^{(M-1)}_s}}{(1+\hbar^{-1} H_{s}^{(M-1)}-\hbar^{-1} H_{t}^{(M)})_{k^{(M-1)}_s}} ~.
\eea
Here we identify $H_{s}^{(I)}$ $(s = 1,...,d_I)$ with Chern roots to the duals of the
universal bundles $\cS_I$:
\be	
0\subset \cS_1 \subset \cS_2 \subset \dots \subset \cS_{M-1} \subset \cS_{M} = 
{\bC}^N \otimes {\cal O}_{\Fl_N}~.
\ee 
Furthermore, the Higgs branch formula of the vortex partition function can be written as
\bea
 Z_{\text{v}}[T^*\Fl(\vec{d})] &=& \sum_{\vec{k}^{(I)}}\prod_{I=1}^{M-1} z_I^{\vert k^{(I)}\vert} \prod_{I=1}^{M-1}\prod_{s\neq t}^{d_I} \tfrac{(1-\hbar^{-1} a_{st}+\hbar^{-1}m)_{k_s^{(I)}-k_t^{(I)}}}{(-\hbar^{-1} a_{st})_{k_s^{(I)}-k_t^{(I)}}} \\
&&\prod_{I=1}^{M-2} \prod_{s=1}^{d_I}\prod_{t=1}^{d_{I+1}} \tfrac{(-\hbar^{-1}a_{st}-\hbar^{-1}m)_{k^{(I)}_s - k^{(I+1)}_t}}{(1-\hbar^{-1}a_{st})_{k^{(I)}_s - k^{(I+1)}_t}}\prod_{s=1}^{d_{M-1}}\prod_{t=1}^{N}\tfrac{(-\hbar^{-1}a_{st}-\hbar^{-1}m)_{k^{(M-1)}_s}}{(1-\hbar^{-1}a_{st})_{k^{(M-1)}_s}} ~.\nonumber
\eea
With the identification $d_I=\sum_{J=1}^I N_J$, this can be regarded as $k_M=0$ specialization of the instanton partition function \eqref{N=2*-general}
\bea
&& Z_{\text{v}}[T^*\Fl(\vec{d})] (z_I,a_i,m,\hbar)\cr
 &&\hspace{1cm}=\sum_{k_1,\cdots,k_{M-1}}  \Big( \prod_{I=1}^{M-1} z_I^{k_I} \Big) \scZ_{\vec{N},k_1,\cdots,k_{M-1}, k_M=0}^{\cN=2^*}(a_i,\mu_{\textrm{adj}}=m-\hbar,\e_1=\hbar)~.
\eea

Among partial flag varieties, the projective space $\CP^{N-1}$ and the Grassmannian $\Gr(r,N)$ play a distinctive role since they are particularly simple. Hence, we write the $J$-functions of their cotangent bundles explicitly. The $J$-function of the cotangent bundle $T^*\CP^{N-1}$ of the projective space is expressed as
\be\label{TCP}
J[T^*\CP^{N-1}]  = \sum_{k}z^{k+\hbar^{-1}H}\frac{(\hbar^{-1} H-\hbar^{-1}m)^N_{k}}{(1+\hbar^{-1} H)^N_{k}}~,
\ee
whereas that of the cotangent bundle $T^*\Gr(r,N)$ of the Grassmannian is given by
\bea\label{TGr}
J[T^*\Gr(r,N)]  &=&  \sum_{\vec{k}}z^{|k_s|+\hbar^{-1}|H_s|} \prod_{s=1}^{r}\frac{(\hbar^{-1} H_{s}-\hbar^{-1}m)^N_{k_s}}{(1+\hbar^{-1} H_{s})^N_{k_s}} \\
&& \prod_{s<t}^r \frac{k_s-k_t+\hbar^{-1}H_s-\hbar^{-1}H_t}{\hbar^{-1}H_s-\hbar^{-1}H_t} \frac{\hbar^{-1}H_s-\hbar^{-1}H_t+1+\hbar^{-1}m}{k_s-k_t+\hbar^{-1}H_s-\hbar^{-1}H_t+1+\hbar^{-1}m} ~.\nonumber
\eea

Hori and Vafa conjectured in \cite{Hori:2000kt} that the $J$-function of the Grassmannian can be obtained by acting Vandermonde differential operators on the product of the $J$-functions of the projective spaces:
\bea
J[\Gr(r,N)](z) = \prod_{s<t}^r \frac{z_s\partial_{z_s}-z_t\partial_{z_t}}{\hbar^{-1}H_{s} - \hbar^{-1}H_{t}} J[{\mathbb{P}}](z_1,\ldots , z_{r}) \Big{\vert}_{z_s=(-1)^{r-1}z}~,
\eea
where we define $ J[{\mathbb{P}}](z_1,\ldots , z_{r}) :=\prod_{s=1}^r J[\CP^{N-1}](z_s)$. This conjecture has been proved in \cite{Bertram:2003}. From the explicit expressions \eqref{TGr} and \eqref{TCP}, it is straightforward to find a similar relation between them
\bea
&&J[T^*\Gr(r,N)](z) \\
&=& \prod_{s<t}^r \frac{z_s\partial_{z_s}-z_t\partial_{z_t}}{\hbar^{-1}H_{s} - \hbar^{-1}H_{t}} \Bigg[  \frac{z_s\partial_{z_s}-z_t\partial_{z_t}+1+\hbar^{-1}m}{\hbar^{-1}H_{s} - \hbar^{-1}H_{t}+1+\hbar^{-1}m}\Bigg]^{-1} J[{\mathbb{T^\ast P}}](z_1,\ldots , z_{r}) \Big{\vert}_{z_s=z}~, \nonumber
\eea
where we define  $J[\mathbb{T^\ast P}](z_1,\ldots , z_{r}) := \prod_{s=1}^r J[T^*\CP^{N-1}](z_s)$. Recently, it was proven in \cite{Ueda:2005,Galkin:2014laa} that the quantum connection of $\Gr(r,N)$ is the $r$-th wedge of the quantum connection of $\CP^{N-1}$. It would be  intriguing to study whether the statement can be extended to their cotangent bundles. Note that the quantum connection of $T^*\CP^{N-1}$ is given by
\bea
\Big[(z\partial_z)^N -z(z\partial_z-\hbar^{-1}m)^N \Big] J[T^*\CP^{N-1}](z)=0~.
\eea

\bibliography{AGT}{}
\bibliographystyle{JHEP}
\end{document}